\documentclass[a4paper,12pt]{article}

\usepackage[pdftex]{graphicx}
\usepackage{amsmath}
\usepackage{amssymb}

\makeatletter
\renewcommand{\section}{\@startsection{section}{1}{0mm}{\baselineskip}{\baselineskip}{\normalsize\bfseries}}
\renewcommand{\subsection}{\@startsection{subsection}{2}{0mm}{\baselineskip}{\baselineskip}{\normalsize\itshape}}
\makeatother

\title{\bf Surface thermodynamics, surface stress, equations at surfaces and triple lines for deformable bodies}

\author{\bf Juan Olives\\
\small CINaM-CNRS\thanks{Associated to Aix-Marseille Universit\'e.}, Campus de Luminy, case 913, 13288 Marseille cedex 9, France\\
\small E-mail: olives@cinam.univ-mrs.fr}

\date{}

\begin{document}

\maketitle

\begin{abstract}

The thermodynamics and mechanics of the surface of a deformable body are studied here, following and refining the general approach of Gibbs. It is first shown that the `local' thermodynamic variables of the state of the surface are only the temperature, the chemical potentials and the surface strain tensor (true thermodynamic variables, for a viscoelastic solid or a viscous fluid). A new definition of the surface stress is given and the corresponding surface thermodynamics equations are presented. The mechanical equilibrium equation at the surface is then obtained. It involves the surface stress and is similar to the Cauchy equation for the volume. Its normal component is a generalization of the Laplace equation. At a (body--fluid--fluid) triple contact line, two equations are obtained, which represent: (i) the equilibrium of the forces (surface stresses) for a triple line fixed on the body; (ii) the equilibrium relative to the motion of the line with respect to the body. This last equation leads to a strong modification of Young's classical capillary equation.

\end{abstract}

\maketitle

\section{Introduction}

\noindent Thermodynamic, chemical and mechanical properties of the surfaces of deformable bodies, with various applications, e.g., on nanostructures, thin films,
adhesions or coating, have been continually investigated, from the early work of
Gibbs \cite{Gibbs:1876-1878} until recent mechanical \cite{Gurtin-Murdoch:1975,Gurtin-Weissmuller-Larche:1998} or thermodynamic \cite{Nozieres-Wolf:1988,Olives:1993,Olives:1996,Sanfeld-Steinchen:2003,Muller-Saul:2004,Rusanov:2005} approaches. However, some basic questions, such as the exact determination of the thermodynamic variables of state of the surface, still persist, as shown, e.g., in the recent expression of the work of deformation of the surface \cite{Nozieres-Wolf:1988,Muller-Saul:2004}. This shows that there is a real need for a rigorous and general thermodynamic approach, in order to determine the thermodynamic variables of state and to obtain the corresponding equations of surface thermodynamics and mechanics. Another problem arises at a (body--fluid--fluid) triple contact line, where the classical elastic theory predicts a singularity of the deformation (namely, an infinite displacement). Generally, authors have assumed some finite thickness of the fluid--fluid interface \cite{Lester:1961,Rusanov:1975}, only considered what occurs outside some neighbourhood of this line \cite{Shanahan-deGennes:1986,Shanahan:1986}, or proposed a new force derived from the volume stresses of the solid \cite{Madasu-Cairncross:2004}. Nevertheless, the validity of Young's classical capillary equation remains problematic and the equilibrium equations at the triple line are in fact completely unknown.

In this paper, the general thermodynamic approach of Gibbs is applied and refined, leading to (i) the determination of the \lq local' thermodynamic variables of state of the surface, (ii) a new definition of the surface stress, (iii) the surface thermodynamics equations, (iv) the equilibrium equations at the surfaces and (v) the equilibrium equations at the triple lines, which imply a strong modification of Young's equation (as in \cite{Olives:1993,Olives:1996}).

\section{Thermodynamic equilibrium conditions}

\begin{figure}[htbp]
\begin{center}
\includegraphics[width=6cm]{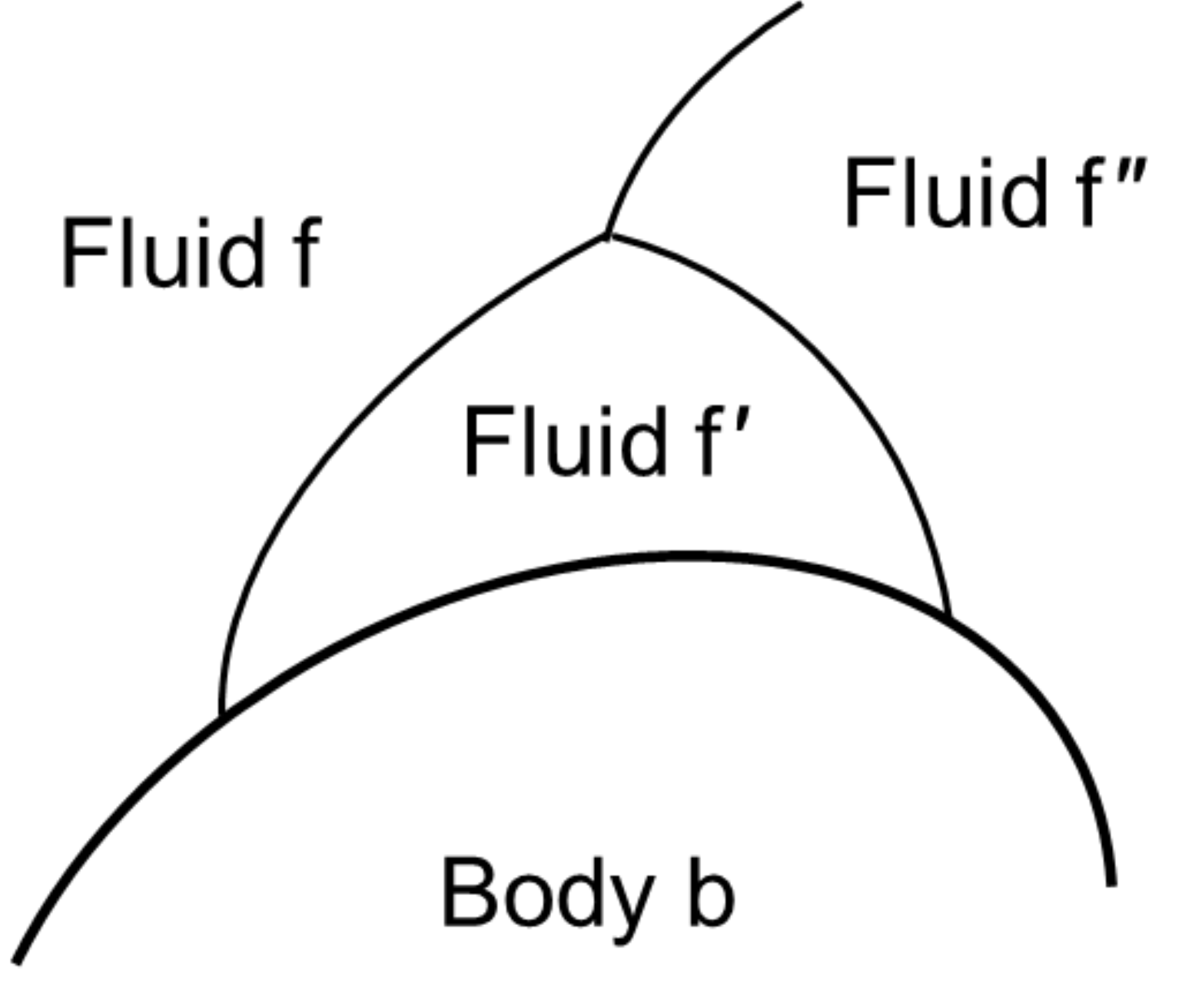}
\end{center}
\caption{The studied system: a deformable body in contact with various fluids.} \label{System}
\end{figure}

\noindent Let us consider a general deformable body $\rm b$ in contact with various immiscible fluids $\rm f$, $\rm f'$,... (figure~\ref{System}). 
No mass exchange is assumed between the body and the fluids, but there may be mass exchanges between all the fluids and the body--fluid ($\rm bf$, $\rm bf'$,...) and fluid--fluid ($\rm ff'$,...) surfaces. For simplicity's sake, we suppose that the body consists of a substance c, the fluids and the fluid--fluid surfaces are composed of the substances 1, 2,..., $n$, with all these components c, 1, 2,..., $n$ being independent. We first recall Gibbs's definition of dividing surfaces, and then define the corresponding `ideal' states and `ideal' transformations.

\subsection{Dividing surfaces, ideal states and ideal transformations}

\noindent In the actual state, a body--fluid interface may be considered as a heterogeneous thin film across which the physical quantities may vary rapidly from their values in the homogeneous body to their values in the homogeneous fluid (`homogeneous' here means \lq with slow and continuous variations of the physical properties'). Following Gibbs \cite{Gibbs:1876-1878}, this actual state is replaced by an ideal state in which the body--fluid interface is replaced by a mathematical dividing surface (without thickness), the body and its physical properties being extrapolated (from their values in the homogeneous body) up to this dividing surface, and the same being made for the fluid on the other side of this surface. For any extensive physical quantity, the surface value of this quantity is then defined as the excess of this physical quantity in the actual state over that of the two homogeneous phases in the ideal state. It is expressed per unit area of the dividing surface, and its value depends on the exact position of this surface, which is arbitrary within (or very close to) the actual interface film. In the present paper, the position of the dividing surface is defined so that the surface mass density of the substance c of the body vanishes. Thus, the body--fluid dividing surfaces only contain (as mass excesses) the substances 1, 2,..., $n$, as in the fluids and the fluid--fluid surfaces. In addition, for each component $i$ (= 1, 2,..., $n$), all the fluid regions and the body--fluid and fluid--fluid surfaces which contain $i$ are supposed to be connected. We also assume that, in the actual interface film, the physical quantities present slow and continuous variations (as those occurring in the homogeneous body or fluid) along the directions parallel to the interface (in contrast to the possible rapid variations of these quantities along the normal to the interface).

\begin{figure}[htbp]
\begin{center}
\includegraphics[width=14cm]{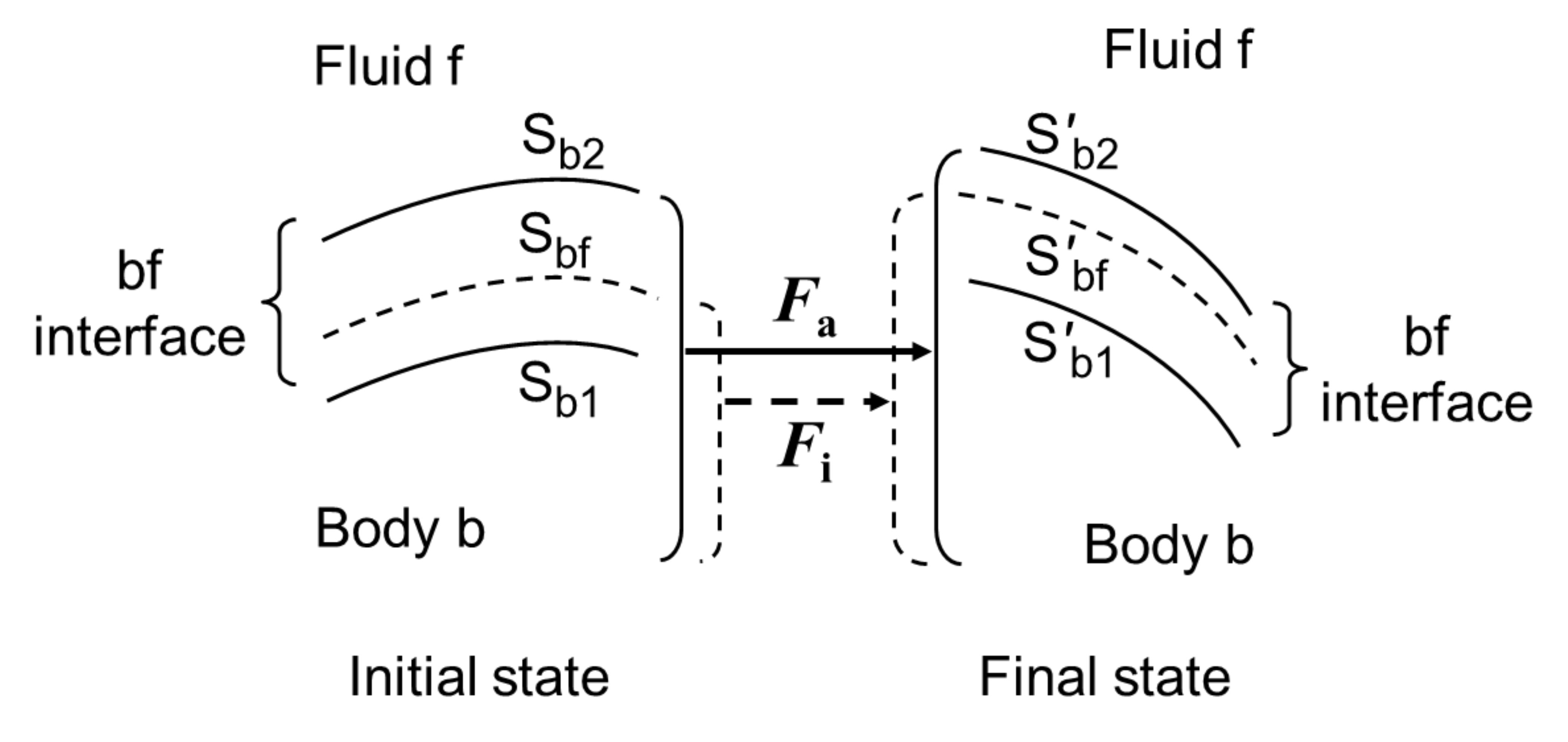}
\end{center}
\caption{The various surfaces ($\rm S_{b1}$, $\rm S_{b2}$ and $\rm S_{bf}$) defined at the body-fluid interface, and the actual and \lq ideal' transformations of the body ($F_{\rm a}$ and $F_{\rm i}$) between the initial and final states (see definitions in the text).} \label{Surfaces}
\end{figure}

Let us now consider two arbitrary states of the body $\rm b$, e.g., in contact with the fluid $\rm f$, and call `initial' and `final' these two states. In the actual initial state, the body is homogeneous up to a surface $\rm S_{b1}$, beyond which the $\rm bf$ interface film takes place (see figure~\ref{Surfaces}). 
Note that a precise definition of $\rm S_{b1}$ is not necessary; in fact, we only need that the body is homogeneous up to $\rm S_{b1}$ and that this surface is close to the interface. On $\rm S_{b1}$, the volume mass density $\rho_{\rm c}$ of the substance c of the body has some value $\rho_{\rm c,b}$. Then, crossing the interface film, $\rho_{\rm c}$ varies rapidly, and finally vanishes in the homogeneous fluid region. We denote $\rm S_{b2}$ the surface such that $\rho_{\rm c}$ is generally $\neq 0$ between $\rm S_{b1}$ and $\rm S_{b2}$, and $\rho_{\rm c} = 0$ beyond $\rm S_{b2}$ (fluid side). The dividing $\rm bf$ surface, as defined above (no excess of mass of the substance c), is denoted $\rm S_{bf}$. In a similar way, the surfaces ${\rm S'_{b1}}$, ${\rm S'_{b2}}$ and $\rm S'_{bf}$ are defined in the final state. Between the initial and final states, there is an actual transformation $F_{\rm a}$, defined in the initial state of the body up to $\rm S_{b2}$, which maps the position $x$ of a material point of the body in the actual initial state, to its position $x'$ in the actual final state. Clearly, ${\rm S'_{b2}} = F_{\rm a}({\rm S_{b2}})$. In addition, $\rm S_{b1}$ and ${\rm S'_{b1}}$ may be chosen so that ${\rm S'_{b1}} = F_{\rm a}({\rm S_{b1}})$. Indeed, such a choice for ${\rm S'_{b1}}$ is justified if the surface $F_{\rm a}({\rm S_{b1}})$ is not situated in the interface film region (in the final state). This last condition can always be satisfied, by choosing---if necessary---a surface $\rm S_{b1}$ slightly separated from the interface film (in the initial state).

If $x_1 \in \rm S_{b1}$ and $x_2 \in \rm S_{b2}$ are two points very close to each other (i.e., $\overrightarrow{x_1 x_2}$ at a small angle from the normal to $\rm S_{b1}$), we assume that $x'_1 = F_{\rm a}(x_1)$ and $x'_2 = F_{\rm a}(x_2)$ will also remain very close to each other. If $y_1 \in \rm S_{b1}$ and $y_2 \in \rm S_{b2}$ are two other such points (thus, $\overrightarrow{x_1 y_1} = \overrightarrow{x_2 y_2}$; this equality and the other ones which follow, have to be considered as good approximations, when the thickness of the interface film is very small), this implies that $\overrightarrow{x'_1 y'_1} = \overrightarrow{x'_2 y'_2}$. By taking infinitesimal vectors $\overrightarrow{x_1 y_1}$, it means that ${\rm D}F_{\rm a}(x_1)|_{\rm P} = {\rm D}F_{\rm a}(x_2)|_{\rm P}$ (restrictions to P of the linear mappings ${\rm D}F_{\rm a}(x_i)$), where ${\rm P} = {\rm T}_{x_1}({\rm S_{b1}}) = {\rm T}_{x_2}({\rm S_{b2}})$ is the plane tangent to the interface. More generally, we assume that ${\rm D}F_{\rm a}(x)|_{\rm P}$ remains unchanged when $x$ crosses the interface (from $x_1$ to $x_2$). 

If the actual initial state is replaced by the ideal one, with respect to the dividing surface $\rm S_{bf}$, an ideal transformation $F_{\rm i}$ may then be defined in the (ideal initial state of the) body up to $\rm S_{bf}$, in the following way: it coincides with $F_{\rm a}$ in the homogeneous part of the body, up to $\rm S_{b1}$, and its derivative ${\rm D}F_{\rm i}$ (linear mapping) in the region between $\rm S_{b1}$ and $\rm S_{bf}$ is extrapolated from the value of the derivative ${\rm D}F_{\rm a}$ in the homogeneous body. Three assertions will now be proved.

1. {\it The surface} $\rm S'_{bf}$ {\it is generally not the image of} $\rm S_{bf}$ {\it by the actual transformation} $F_{\rm a}$. This may be shown with the following simple example. Let the interface be plane, $x$ the normal axis, $x = 0$ and $x = d$ the respective positions of $\rm S_{b1}$ and $\rm S_{b2}$, and $\rho_{\rm c}$ a linear decreasing function of $x$, with values $\rho_{\rm c,b}$ at $x = 0$ and 0 at $x = d$, i.e. $\rho_{\rm c} = \rho_{\rm c,b}(1-\xi)$ with $0\leq \xi = x/d \leq 1$, which leads to the position $x = d/2$ of $\rm S_{bf}$. Let the actual transformation $F_{\rm a}: x \rightarrow x'$ be a simple compression along the $x$ axis, such that $d x'/d x$ linearly decreases from the value 1 at $\xi = 0$ to the small value $\varepsilon > 0$ at $\xi = 1-\varepsilon$ (i.e. $d x'/d x = 1-\xi$), and then remains constant ($= \varepsilon$) for $1-\varepsilon \leq \xi \leq 1$. Under this transformation, the volume mass density in the actual final state will be
\begin{align*}
\rho'_{\rm c} = \frac{\rho_{\rm c}}{\frac{d x'}{d x}}&=\rho_{\rm c,b}\quad\textrm{for}\; 0\leq \xi \leq 1-\varepsilon\\
&=\rho_{\rm c,b} \frac{1-\xi}{\varepsilon}\quad\textrm{for}\; 1-\varepsilon \leq \xi \leq 1.
\end{align*}
Since $\rho'_{\rm c}$ is constant for $0\leq \xi \leq 1-\varepsilon$, the dividing surface $\rm S'_{bf}$ in the final state will be situated at $x'$ such that $1-\varepsilon < \xi = x/d < 1$, i.e. $\xi$ very close to 1. The surface $\rm S_{bf}$ being situated at $\xi = 1/2$, this shows that $\rm S'_{bf}$ cannot be the image of $\rm S_{bf}$ by $F_{\rm a}$.

\begin{figure}[htbp]
\begin{center}
\includegraphics[width=10cm]{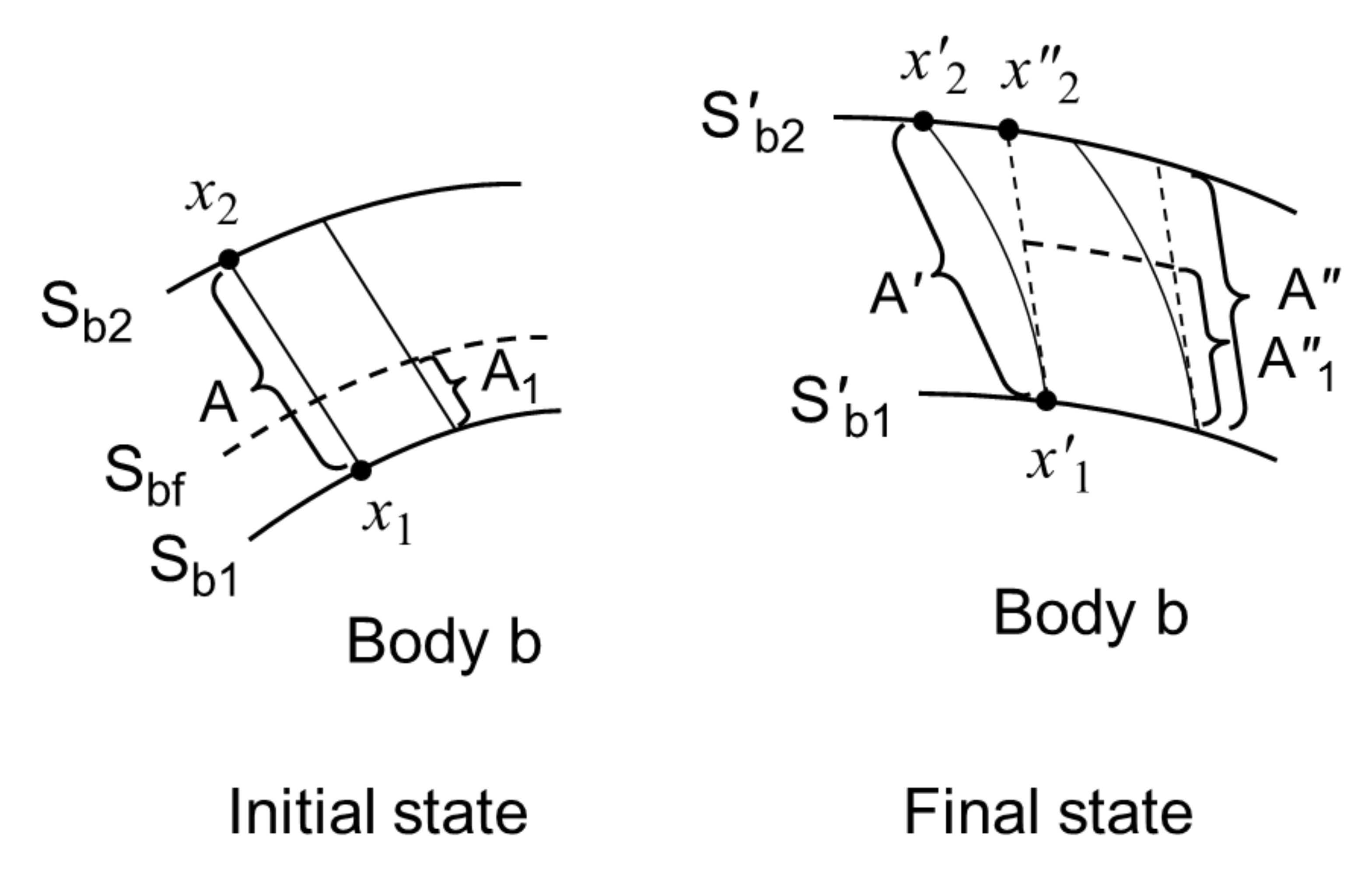}
\end{center}
\caption{The image of $\rm S_{bf}$ by the ideal transformation $F_{\rm i}$ is equal to $\rm S'_{bf}$ (see proof, in the text).} \label{Ideal_transformation}
\end{figure}

2. {\it The surface} $\rm S'_{bf}$ {\it is the image of} $\rm S_{bf}$ {\it by the ideal transformation} $F_{\rm i}$. In order to prove this assertion, let us consider, in the actual initial state, $x_1 \in \rm S_{b1}$, $x_2 \in \rm S_{b2}$ (with $\overrightarrow{x_1 x_2}$ at a small angle from the normal to $\rm S_{b1}$), a small parallelepiped $\rm A$ with edge $x_1 x_2$, limited by $\rm S_{b1}$ and $\rm S_{b2}$, and a similar parallelepiped $\rm A_1$ limited by $\rm S_{b1}$ and $\rm S_{bf}$ (figure~\ref{Ideal_transformation}).
In the final state, $\rm A$ is transformed into ${\rm A'} = F_{\rm a}(\rm A)$, which is not a parallelepiped, since its (curved) \lq edge' $x'_1 x'_2$ suffered an heterogeneous deformation. Nevertheless, all the sections of ${\rm A'}$ by the planes parallel to ${\rm P'} = {\rm T}_{x'_1}({\rm S'_{b1}})$ represent the same parallelogram (up to a translation), since ${\rm D}F_{\rm a}(x)|_{\rm P}$ remains unchanged when $x$ moves from $x_1$ to $x_2$ (as noted above; ${\rm P} = {\rm T}_{x_1}({\rm S_{b1}})$). Let us consider the straight line $x'_1 x''_2$ (with $x''_2 \in \rm S'_{b2}$) tangent to the curved edge $x'_1 x'_2$ at $x'_1$, and the parallelepiped $\rm A''$ with edge $x'_1 x''_2$, having the same face on $\rm S'_{b1}$ as $\rm A'$ (see figure~\ref{Ideal_transformation}). By the definition of $F_{\rm i}$, ${\rm D}F_{\rm i}(x)$ is constant and equal to ${\rm D}F_{\rm a}(x_1)$, between $\rm S_{b1}$ and $\rm S_{bf}$, in $\rm A_1$, which implies that ${\rm A''_1} = F_{\rm i}(\rm A_1)$ is a parallelepiped which, like $\rm A''$, rests on the edge $x'_1 x''_2$ and has the same face on $\rm S'_{b1}$ as $\rm A'$.

By the definition of $\rm S_{bf}$ (no excess of mass of the constituent c), the mass of c contained in $\rm A$ (in the actual state) is equal to the mass of c contained in $\rm A_1$ in the ideal state (i.e., if $\rm A_1$ is filled with only the substance c, with a constant density $\rho_{\rm c}$, equal to $\rho_{\rm c,b} = \rho_{\rm c}(x_1)$). Note that this property does not depend on the choice of the parallelepiped $\rm A$ (i.e., its face on $\rm S_{b1}$ and the orientation of $\overrightarrow{x_1 x_2}$).

Under the transformation $F_{\rm a}$, the mass of c contained in $\rm A$ is equal to that contained in $\rm A'$ (for the actual states). In addition, if $\rm A'$ is `cut up' into very thin slices parallel to $\rm P'$, then, by appropriate sliding of these slices (with respect to each other), we can obtain the parallelepiped $\rm A''$, which shows that the mass of c contained in $\rm A'$ is equal to that contained in $\rm A''$ (for the actual state).

Now, let us transport by $F_{\rm i}$ the substance c contained in $\rm A_1$ in the ideal state: ${\rm A''_1} = F_{\rm i}(\rm A_1)$ will thus contain the same mass of c as that contained in $\rm A_1$ (in the ideal state). In addition, the density $\rho'_{\rm c}$ will be constant in ${\rm A''_1}$, and equal to $\rho'_{\rm c}(x'_1)$ (since $\rho_{\rm c}$ and ${\rm D}F_{\rm i}$ are constant in $\rm A_1$, and respectively equal to $\rho_{\rm c}(x_1)$ and ${\rm D}F_{\rm a}(x_1)$). The various preceding mass equalities finally lead to: the mass of c contained in ${\rm A''_1}$ (filled with only the substance c, with a constant density $\rho'_{\rm c}$, equal to $\rho'_{\rm c}(x'_1)$) is equal to that contained in $\rm A''$ in the actual state. This exactly means that the parallelepiped ${\rm A''_1}$ is limited by the surface $\rm S'_{bf}$ (defined by: no excess of mass of the constituent c), i.e., that ${\rm S'_{bf}} = F_{\rm i}(\rm S_{bf})$. We may also say that the mass density in ${\rm A''_1}$ (resulting from the $F_{\rm i}$ transport) is that of the ideal final state (as defined at the beginning of the section).

3. {\it The volume mass density of the substance} c {\it of the body in the ideal final state results from the transport by the ideal transformation} $F_{\rm i}$ {\it of the mass density of this substance in the ideal initial state}. This is a consequence of the previous last conclusion.

Let us now consider a third arbitrary state of the body $\rm b$, called `reference state', and define, as above for the initial and final states, the corresponding surfaces $\rm S_{0,b1}$, $\rm S_{0,b2}$ and $\rm S_{0,bf}$ (subscript 0 for the reference state). The actual and ideal transformations are respectively denoted $F_{0,\rm a}$ and $F_{0,\rm i}$, between the reference and initial states, and $F'_{0,\rm a}$ and $F'_{0,\rm i}$, between the reference and final states. We have the new assertion:

4. {\it The ideal transformation} $F'_{0,\rm i}$ {\it between the reference and final states is equal to the ideal transformation} $F_{0,\rm i}$ {\it between the reference and initial states composed with the ideal transformation} $F_{\rm i}$ {\it between the initial and final states:} $F'_{0,\rm i} = F_{\rm i} \circ F_{0,\rm i}$. Indeed, let $x_0$ be situated between $\rm S_{0,b1}$ and $\rm S_{0,bf}$, $x_{0,1} \in \rm S_{0,b1}$ be very close to $x_0$, and denote $x = F_{0,\rm i}(x_0)$, $x_1 = F_{0,\rm a}(x_{0,1})$, $x' = F_{\rm i}(x)$, $x'_1 = F_{\rm a}(x_1) = F'_{0,\rm a}(x_{0,1})$. Then, $\overrightarrow{x'_1 x'} = {\rm D}F_{\rm a}(x_1) \cdot \overrightarrow{x_1 x} = {\rm D}F_{\rm a}(x_1) \cdot {\rm D}F_{0,\rm a}(x_{0,1}) \cdot \overrightarrow{x_{0,1} x_0} = {\rm D}F'_{0,\rm a}(x_{0,1}) \cdot \overrightarrow{x_{0,1} x_0}$, hence $x' = F'_{0,\rm i}(x_0)$, i.e. $(F_{\rm i} \circ F_{0,\rm i})(x_0) = F'_{0,\rm i}(x_0)$.

\subsection{Thermodynamic equilibrium conditions}

\noindent We consider that the system is closed and bounded by a closed surface $\Sigma$, and write the Gibbs equilibrium criterion
\begin{gather*}
\delta U = \delta W_{\rm e}
\end{gather*}
($U$ is the internal energy and $\delta W_{\rm e}$ the work of the external forces), for all variations $\delta$ of the system such that the entropy $S$ and the masses $m_i$ of the various components ($i$ = 1,..., $n$) are constant. If the bounding surface $\Sigma$, the points of the body which belong to this surface, and the lines in which the fluid--fluid surfaces meet this surface, are all fixed, then $\delta W_{\rm e}$ is reduced to the work of the forces of gravity (there is no work of the fluid pressures on $\Sigma$, since they are normal to $\Sigma$ and the fluid displacements are tangent to $\Sigma$), and the criterion may be written
\begin{gather*}
\delta (U+V) = 0,
\end{gather*}
where $V$ is the potential energy of gravity. Note that, in this variation, we exclude the formation of new fluids or new (body--fluid or fluid--fluid) surfaces.

We thus have two states of the system: the present state and the (arbitrary) varied state, i.e., after the variation $\delta$. Let us respectively call these two states `initial' and `final' and then apply the definitions and results of the preceding section. We may thus write the internal energy of the actual present state as
\begin{align*}
U &= U_{\rm b} + \sum_{\rm f} U_{\rm f} + \sum_{\rm ff'} U_{\rm ff'} + \sum_{\rm bf} U_{\rm bf}\\
&= \int_{\rm b} d U + \sum_{\rm f} \int_{\rm f} d U + \sum_{\rm ff'} \int_{\rm ff'} d U + \sum_{\rm bf} \int_{\rm bf} d U, 
\end{align*}
where $d$ refers to an element of volume or surface, $U_{\rm b}$ and $U_{\rm f}$ are the energies of the homogeneous phases in the ideal state (i.e., extrapolated up to the dividing surfaces), and $U_{\rm ff'}$ and $U_{\rm bf}$ the excesses of the energy for the corresponding dividing surfaces ($\rm S_{ff'}$ as defined by Gibbs; $\rm S_{bf}$ as defined above). Note that, in the present approach, we only take into account the volume and the surface effects (as a first approximation, intrinsic line effects, such as line energy, line tension,... are here neglected). Similarly, for the varied state,
\begin{align*}
U' &= U'_{\rm b} + \sum_{\rm f} U'_{\rm f} + \sum_{\rm ff'} U'_{\rm ff'} + \sum_{\rm bf} U'_{\rm bf}\\
&= \int_{\rm b} d U' + \sum_{\rm f} \int_{\rm f} d U' + \sum_{\rm ff'} \int_{\rm ff'} d U' + \sum_{\rm bf} \int_{\rm bf} d U', 
\end{align*}
with respect to the dividing surfaces of this state, i.e., $\rm S'_{ff'}$ (as defined by Gibbs) and $\rm S'_{bf}$ (as defined above). In the difference $\delta U = U' - U$, the first term is 
\begin{gather*}
\delta \int_{\rm b} d U = \int_{\rm b} d U' - \int_{\rm b} d U.
\end{gather*}
In the last integral, we may consider that the body $\rm b$, in its ideal present state (up to $\rm S_{bf}$), has been divided into infinitesimal elements of volume (of energy $d U$). The assertions 2 and 3 of the preceding section show that the body $\rm b$, in its ideal varied state (up to $\rm S'_{bf}$), may be exactly divided into the elements which are the images of the preceding ones by the ideal transformation $F_{\rm i}$ (from the ideal present state to the ideal varied state). If $\delta d U = d U' - d U$ refers to an element of the ideal present state and its corresponding image by $F_{\rm i}$, we may then write
\begin{gather*}
\delta \int_{\rm b} d U = \int_{\rm b} \delta d U,
\end{gather*}
and since each element is a closed `system' (with respect to the geometrical transformation $F_{\rm i}$), for any infinitesimal reversible thermodynamic transformation $\delta$:
\begin{gather*}
\delta \int_{\rm b} d U = \int_{\rm b} (T\,\delta d S + \sigma \cdot \delta \varepsilon\,d v)
\end{gather*}
(in the ideal states; $\delta$ refers to $F_{\rm i}$; all the quantities are extrapolated up to the dividing surfaces), where $T$ is the temperature, $v$ the volume, $\sigma$ the stress tensor at equilibrium and $\delta \varepsilon$ the strain tensor corresponding to $F_{\rm i}$. According to assertion 4 of the preceding section, this may also be written in the Lagrangian representation, with respect to a given reference state of the body:
\begin{gather*}
\delta \int_{\rm b} d U = \int_{\rm b} (T\,\delta d S + \pi \cdot \delta e\,d v_0),
\end{gather*}
where $v_0$ is the volume in the reference state, $\pi$ the Piola--Kirchhoff stress tensor (relative to the reference state) at equilibrium, and $e$ the Green--Lagrange strain tensor (relative to the reference state, i.e., associated to $F_{0,\rm i}$; $\delta e = e(F'_{0,\rm i}) - e(F_{0,\rm i})$).

For the $\rm f$ and $\rm ff'$ terms in $\delta U = U' - U$, the expressions of Gibbs are used:
\begin{align}
\delta \int_{\rm f} d U &= \int_{\rm f} (T\,\delta d S - p\,\delta d v + \sum_i \mu_i\,\delta d m_i)\nonumber\\
\delta \int_{\rm ff'} d U &= \int_{\rm ff'} (T\,\delta d S + \gamma\,\delta d a + \sum_i \mu_i\,\delta d m_i)\label{ff'internalenergy}
\end{align}
(in the ideal states; in the integrals on $\rm f$, the quantities are extrapolated up to the dividing surfaces; the $\rm ff'$ integrals refer to the dividing surfaces, $d U$, $d S$ and $d m_i$ being surface excesses and $d a$ an element of area), where $p$ is the fluid pressure, $\gamma$ the fluid--fluid surface tension and $\mu_i$ the chemical potential per unit mass of component $i$. Note that, in this case, each element of volume or surface is treated as an open `system'.

The above arguments may similarly be applied to the potential energy of gravity:
\begin{align*}
V &= V_{\rm b} + \sum_{\rm f} V_{\rm f} + \sum_{\rm ff'} V_{\rm ff'} + \sum_{\rm bf} V_{\rm bf}\\
&= \int_{\rm b} g\,z\,d m + \sum_{\rm f} \int_{\rm f} g\,z\,d m
+ \sum_{\rm ff'} \int_{\rm ff'} g\,z\,d m + \sum_{\rm bf} \int_{\rm bf} g\,z\,d m, 
\end{align*}
where $V_{\rm b}$ and $V_{\rm f}$ correspond to the homogeneous phases in the ideal state (i.e., extrapolated up to the dividing surfaces: $\rm S_{ff'}$, $\rm S_{bf}$), $V_{\rm ff'}$ and $V_{\rm bf}$ are the excess quantities for these dividing surfaces, $g$ is the gravity field (supposed constant), $z$ the height and $m$ the mass. In the $\rm ff'$ and $\rm bf$ integrals, $d m$ represents the surface excess for the dividing surfaces (the excess of $\int_{\rm vol.} g\,z\,d m$ is equal to $\int_{\rm surf.} g\,z\,(\textrm{excess of }d m)$, since, crossing the interface film, $z$ remains almost constant whereas the volume mass density varies rapidly). Then,
\begin{gather*}
\delta \int_{\rm b} g\,z\,d m = \int_{\rm b} \delta(g\,z\,d m) = \int_{\rm b} g\,\delta z\,d m,
\end{gather*}
each element being a closed `system' with respect to the transformation $F_{\rm i}$ ($\delta$ refers to $F_{\rm i}$), and
\begin{align*}
\delta \int_{\rm f} g\,z\,d m &= \int_{\rm f} (g\,\delta z\,d m + g\,z\,\delta d m)\\
\delta \int_{\rm ff'} g\,z\,d m &= \int_{\rm ff'} (g\,\delta z\,d m + g\,z\,\delta d m)
\end{align*}
(in the ideal states; the elements of volume or surface are here treated as open `systems').

Finally, the Gibbs equilibrium criterion may be written
\begin{align}
&\int_{\rm b} (T\,\delta d S + \pi \cdot \delta e\,d v_0) 
+ \int_{\rm b} g\,\delta z\,d m \nonumber\\
&+ \sum_{\rm f} \int_{\rm f} (T\,\delta d S - p\,\delta d v + \sum_i \mu_i\,\delta d m_i)
+ \sum_{\rm f} \int_{\rm f} (g\,\delta z\,d m + g\,z\,\delta d m) \nonumber\\
&+ \sum_{\rm ff'} \int_{\rm ff'} (T\,\delta d S + \gamma\,\delta d a + \sum_i \mu_i\,\delta d m_i)
+ \sum_{\rm ff'} \int_{\rm ff'} (g\,\delta z\,d m + g\,z\,\delta d m) \nonumber\\
&+ \sum_{\rm bf} \delta \int_{\rm bf} d U + \sum_{\rm bf} \delta \int_{\rm bf} g\,z\,d m = 0, \label{variationalproblem}
\end{align}
and the associated conditions of constant $S$ and $m_i$ are
\begin{gather}
\int_{\rm b} \delta d S + \sum_{\rm f} \int_{\rm f} \delta d S 
+ \sum_{\rm ff'} \int_{\rm ff'} \delta d S + \sum_{\rm bf} \delta \int_{\rm bf} d S = 0 \label{constantS}\\
\sum_{\rm f} \int_{\rm f} \delta d m_i + \sum_{\rm ff'} \int_{\rm ff'} \delta d m_i 
+ \sum_{\rm bf} \delta \int_{\rm bf} d m_i = 0\quad(i = 1,...,n).\label{constantmi}
\end{gather}
Note that these expressions, and all the following ones, refer to the ideal states, which means that all the quantities of the volume integrals are extrapolated up to the dividing surfaces (and, in the $\rm b$ integrals, $\delta$ refers to $F_{\rm i}$), and the surface integrals refer to these dividing surfaces ($d S$, $d U$, $d m$ and $d m_i$ being surface excesses).

From this point, we may follow the proof given in \cite{Olives:1993} (extension of the approach of Gibbs \cite{Gibbs:1876-1878}), here applied to the case of a general deformable body, which first shows that the equilibrium (\ref{variationalproblem})--(\ref{constantmi}) of the system is equivalent to
 
(i) the thermal and chemical equilibrium equations:
\begin{gather}
T = \textrm{constant in space} \label{Tequilibrium}\\
\mu_i + g\,z = M_i = \textrm{constant in space} \quad(i = 1,...,n)\label{muiequilibrium}
\end{gather}

(ii) the mechanical equilibrium equations concerning only the system of the fluids:

in each fluid $\rm f$
\begin{gather}
\frac{d p}{d z} = -\rho\,g\label{pequilibrium}
\end{gather}

in each $\rm ff'$ surface
\begin{gather}
\frac{d \gamma}{d z} = \rho_{\rm s}\,g\label{gammaequilibrium}\\
p - p' = \gamma\,(\frac{1}{R_1} + \frac{1}{R_2}) + \rho_{\rm s}\,g \cos\theta\label{pp'equilibrium}
\end{gather}

in each $\rm ff'f''$ triple line
\begin{gather}
\sum_{\rm ff'} \gamma\,\nu = 0\label{gammanuequilibrium}
\end{gather}
($\rho$ is the mass per unit volume, $\rho_{\rm s}$ the mass per unit area (surface excess); $p$ refers to $\rm f$, $p'$ to $\rm f'$; $R_1$ and $R_2$ are the principal curvature radii of the $\rm ff'$ surface, considered positive when the centres are on the $\rm f$ side; $\theta$ is the angle between the O$z$ axis and the normal to the $\rm ff'$ surface, oriented from $\rm f$ to $\rm f'$; at the $\rm ff'f''$ line, and for each $\rm ff'$ surface, $\nu$ is the unit vector normal to the $\rm ff'f''$ line, tangent to the $\rm ff'$ surface, and oriented from the line to the interior of $\rm ff'$; in (\ref{pequilibrium}) and (\ref{gammaequilibrium}), $p$, $\gamma$, $\rho$ and $\rho_{\rm s}$ only depend on $z$; all these equations (\ref{Tequilibrium})--(\ref{gammanuequilibrium}) were written by Gibbs \cite{Gibbs:1876-1878}) and

\begin{figure}[htbp]
\begin{center}
\includegraphics[width=13cm]{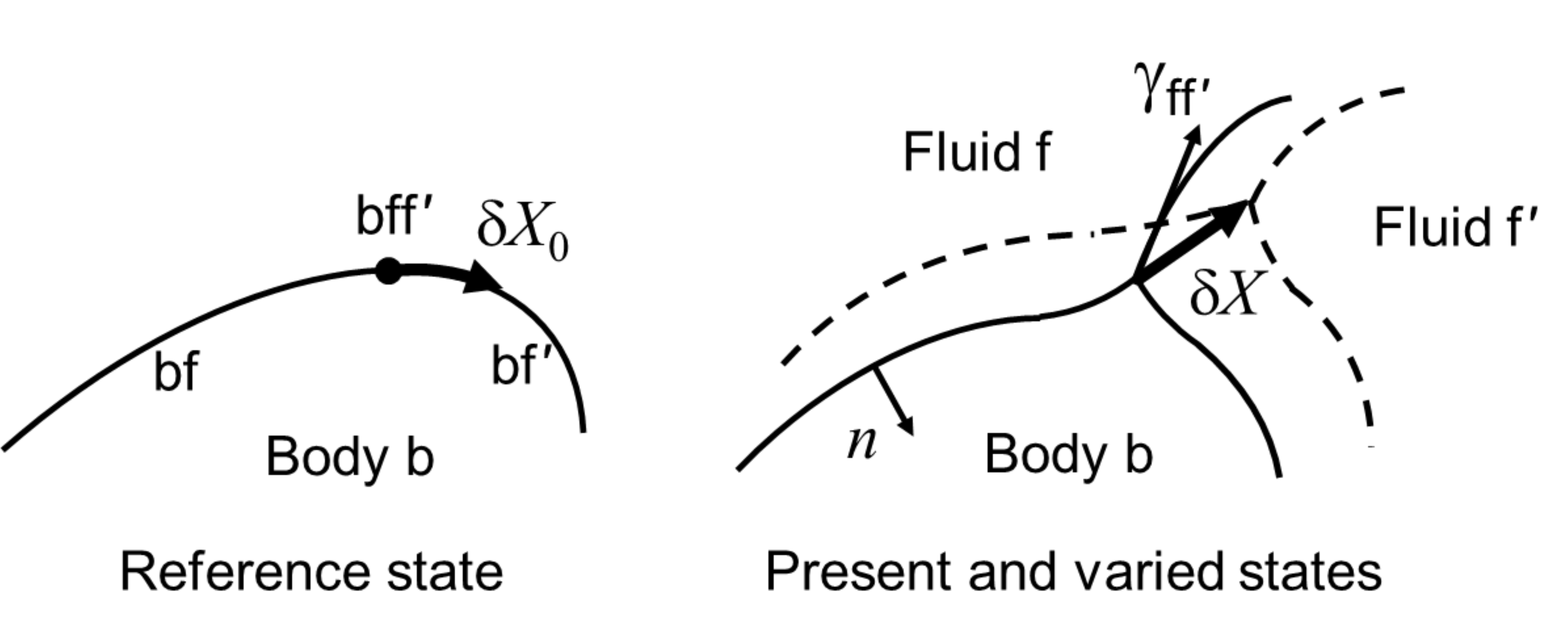}
\end{center}
\caption{Displacement $\delta X$ of the $\rm bff'$ triple line in space, between the present and the varied states, and displacement $\delta X_0$ of this line with respect to the body, in the reference state.} \label{Line}
\end{figure}

(iii) the following new mechanical equilibrium condition (of variational form) which concerns only the body, the body--fluid surfaces and the body--fluid--fluid lines:
\begin{align}
&\int_{\rm b} \pi \cdot \delta e\,d v_0 
- \int_{\rm b} \rho\,\bar g \cdot \delta x\,d v \nonumber\\
&- \sum_{\rm bf} \int_{\rm bf} p\,n \cdot \delta x\,d a
- \sum_{\rm bff'} \int_{\rm bff'} \gamma_{\rm ff'}\,\nu_{\rm ff'}\cdot \delta X\,d l \nonumber\\
&+ \sum_{\rm bf} (\delta\int_{\rm bf} d U -T\,\delta\int_{\rm bf} d S -\sum_i M_i\,\delta\int_{\rm bf} d m_i + \delta\int_{\rm bf} g\,z\,d m) = 0, \label{bodyvariational0}
\end{align}
where $\bar g$ is the (constant) gravity vector field, $\delta x$ the displacement of a material point of the body (by $F_{\rm i}$, up to the dividing surfaces $\rm S_{bf}$), $n$ the unit vector normal to the $\rm bf$ surface, oriented from $\rm f$ to $\rm b$ ($n \cdot \delta x$ also represents the normal displacement of the surface, from $\rm S_{bf}$ to $\rm S'_{bf}$, positively measured from $\rm f$ to $\rm b$), $\nu_{\rm ff'}$ the unit vector normal to the $\rm bff'$ line, tangent to the $\rm ff'$ surface, and oriented from the line to the interior of $\rm ff'$, $d l$ the length of a $\rm bff'$ line element, and $\delta X$ the (vector) displacement of the $\rm bff'$ line, perpendicular to the line (figure~\ref{Line}). 
Note that (\ref{pequilibrium}) and (\ref{gammaequilibrium}) are consequences of (\ref{Tequilibrium}) and (\ref{muiequilibrium}), since $p$ and $\gamma$ are functions of $T$ and $\mu_i$ which satisfy \cite{Gibbs:1876-1878}
\begin{align}
d p &= S_v\,d T + \sum_i m_{i,v}\,d \mu_i\nonumber\\
d \gamma &= -S_a\,d T - \sum_i m_{i,a}\,d \mu_i\label{gammaGibbs}
\end{align}
(the subscripts $v$ and $a$ respectively denote the volume density and the surface density).

The terms of the last line of (\ref{bodyvariational0}) may be calculated in the (ideal) reference state of the body (i.e., on $\rm S_{0,bf}$). With the help of the above assertions 2 (${\rm S'_{bf}} = F_{\rm i}(\rm S_{bf})$, ${\rm S_{bf}} = F_{0,\rm i}(\rm S_{0,bf})$ and ${\rm S'_{bf}} = F'_{0,\rm i}(\rm S_{0,bf})$) and 4 ($F'_{0,\rm i} = F_{\rm i} \circ F_{0,\rm i}$), we may write, e.g., for the first term
\begin{gather*}
\delta \int_{\rm bf} d U = \delta \int_{\rm bf} U_{a_0}\,d a_0
= \int_{\rm bf} \delta U_{a_0}\,d a_0 + \sum_{\rm f'} \int_{\rm bff'} U_{a_0,\rm bf}\,\delta X_0\,d l_0,
\end{gather*}
where the subscript 0 is used for the reference state, $U_{a_0} = d U/d a_0$, $\delta U_{a_0}$ is the variation at a fixed point of $\rm S_{0,bf}$, and $\delta X_0$ the (scalar) displacement of the $\rm bff'$ line, measured in the reference state, perpendicular to that line in the reference state, and positively considered from $\rm bf$ to $\rm bf'$ (figure~\ref{Line}). Then,
\begin{gather*}
\sum_{\rm bf} \delta \int_{\rm bf} d U
= \sum_{\rm bf} \int_{\rm bf} \delta U_{a_0}\,d a_0 
+ \sum_{\rm bff'} \int_{\rm bff'} (U_{a_0,\rm bf} - U_{a_0,\rm bf'})\,\delta X_0\,d l_0.
\end{gather*}

The two last terms of the last line of (\ref{bodyvariational0}) only depend on the variations of the masses and the geometrical positions. In particular, they do not depend on $\delta M_i$ (or $\delta \mu_i$), and we may thus use, e.g., $\delta M_i = 0$ (i.e., $\delta \mu_i = - g\,\delta z$) to calculate these terms:
\begin{align*}
-\sum_i M_i\,\delta \int_{\rm bf} d m_i + \delta \int_{\rm bf} g\,z\,d m
&= \delta \int_{\rm bf} (-\sum_i M_i\,d m_i + g\,z\,d m)\\
&= -\sum_i \delta \int_{\rm bf} \mu_i\,d m_i
= -\sum_i \delta \int_{\rm bf} \mu_i\,m_{i,a_0}\,d a_0.
\end{align*}
By applying the same method as above (with $\mu_i\,m_{i,a_0}$ instead of $U_{a_0}$), and according to $\delta \mu_i = - g\,\delta z$, we finally obtain for the last line of (\ref{bodyvariational0}):
\begin{align*}
&\sum_{\rm bf} (\delta \int_{\rm bf} d U -T\,\delta \int_{\rm bf} d S -\sum_i M_i\,\delta \int_{\rm bf} d m_i 
+ \delta \int_{\rm bf} g\,z\,d m)\\
&= \sum_{\rm bf} \int_{\rm bf} g\,\delta z\,d m
+ \sum_{\rm bf} \int_{\rm bf} (\delta U_{a_0} - T\,\delta S_{a_0} - \sum_i \mu_i\,\delta m_{i,a_0})\,d a_0\\
&+ \sum_{\rm bff'} \int_{\rm bff'} (\gamma_{0,\rm bf} - \gamma_{0,\rm bf'})\,\delta X_0\,d l_0,
\end{align*}
where $\gamma_0 = U_{a_0} - T\,S_{a_0} - \sum_i \mu_i\,m_{i,a_0}$ is the excess of grand potential (on $\rm S_{bf}$, for $\gamma_{0,\rm bf}$), per unit area in the reference state (on $\rm S_{0,bf}$). The equilibrium condition (\ref{bodyvariational0}) may then be written as
\begin{align}
&\int_{\rm b} \pi \cdot \delta e\,d v_0 
- \int_{\rm b} \rho\,\bar g \cdot \delta x\,d v \nonumber\\
&- \sum_{\rm bf} \int_{\rm bf} p\,n \cdot \delta x\,d a
- \sum_{\rm bf} \int_{\rm bf} \rho_{\rm s}\,\bar g \cdot \delta x\,d a
+ \sum_{\rm bf} \int_{\rm bf} (\delta U_{a_0} - T\,\delta S_{a_0} 
- \sum_i \mu_i\,\delta m_{i,a_0})\,d a_0 \nonumber\\
&- \sum_{\rm bff'} \int_{\rm bff'} \gamma_{\rm ff'}\,\nu_{\rm ff'}\cdot \delta X\,d l
+ \sum_{\rm bff'} \int_{\rm bff'} (\gamma_{0,{\rm bf}} - \gamma_{0,{\rm bf'}}) 
\,\delta X_0\,d l_0 = 0, \label{bodyvariational1}
\end{align}
in which $\delta$ is an arbitrary variation such that, on the surface $\Sigma$ which bounds the system, the points of the body and the points of the body--fluid--fluid lines remain fixed. Note that $\delta x$ is the displacement by $F_{\rm i}$ (up to the dividing surfaces $\rm S_{bf}$).

\section{Surface thermodynamics and surface stress}

\noindent Now, let us take a bounding surface $\Sigma$ which only encloses one fluid $\rm f$ and the body $\rm b$. The $\rm S_{bf}$ surface enclosed in $\Sigma$ is bounded by the closed curve $\Gamma = \rm S_{bf} \cap \Sigma$. The above equilibrium condition (\ref{bodyvariational1}) takes the form:
\begin{align}
&\int_{\rm b} \pi \cdot \delta e\,d v_0 
- \int_{\rm b} \rho\,\bar g \cdot \delta x\,d v \nonumber\\
&- \int_{\rm bf} p\,n \cdot \delta x\,d a
- \int_{\rm bf} \rho_{\rm s}\,\bar g \cdot \delta x\,d a
+ \int_{\rm bf} (\delta U_{a_0} - T\,\delta S_{a_0} 
- \sum_i \mu_i\,\delta m_{i,a_0})\,d a_0 = 0, \label{bfvariational1}
\end{align}
for any variation such that the points of the body which belong to $\Sigma$ remain fixed (by the actual transformation $F_{\rm a}$; then, also, by the ideal transformation $F_{\rm i}$, since $F_{\rm i}$ is the extrapolation of $F_{\rm a}$, between $\rm S_{b1}$ and $\rm S_{bf}$); $\rm b$ and $\rm bf$ respectively denote the parts of the body and the $\rm S_{bf}$ surface enclosed in $\Sigma$. Let us use the usual Eulerian representations of stress ($\sigma$) and strain ($\delta \varepsilon = \frac{1}{2} ({\rm D}w +\,({\rm D}w)^*)$, where $w = \delta x$), and Green's formula:
\begin{align}
&\int_{\rm b} \pi \cdot \delta e\,d v_0 
= \int_{\rm b} \sigma \cdot \delta \varepsilon\,d v
= \int_{\rm b} \frac{\sigma + \sigma^*}{2}\cdot \delta \varepsilon\,d v 
= \int_{\rm b} \frac{\sigma + \sigma^*}{2}\cdot {\rm D}w\,d v \nonumber\\
&= \int_{\rm b} \sigma \cdot {\rm D}w\,d v
- \int_{\rm b} \frac{\sigma - \sigma^*}{2}\cdot {\rm D}w\,d v \nonumber\\
&= - \int_{\rm b} ({\rm div}\,\sigma) \cdot w\,d v
- \int_{\rm bf} (\sigma \cdot n) \cdot w\,d a
- \int_{\rm b} \frac{\sigma - \sigma^*}{2}\cdot {\rm D}w\,d v.\label{divergencesigma}
\end{align}
The preceding equilibrium condition is then equivalent to

(i) the mechanical equilibrium equations of the body
\begin{gather}
{\rm div}\,\sigma + \rho\,\bar g = 0\label{bequilibrium1}\\
\sigma^* = \sigma\label{bequilibrium2} 
\end{gather}
(obtained by fixing the points of the $\rm S_{bf}$ surface, and their thermodynamic state, in (\ref{bfvariational1}); note that, with the presence of volume couples of forces, of moment $M\,d v$ ($M$ written as an antisymmetric tensor), the new term $-\int_{\rm b} \frac{1}{2} M\cdot {\rm D}w\,d v$ would appear in (\ref{bfvariational1}), and (\ref{bequilibrium2}) would become $\sigma - \sigma^* + M$ = 0; in our case, $M = 0$, and $\sigma$ and $\pi$ are symmetric) and

(ii) the following mechanical equilibrium condition for $\rm S_{bf}$:
\begin{align}
&- \int_{\rm bf} (\sigma \cdot n) \cdot \delta x\,d a 
- \int_{\rm bf} p\,n \cdot \delta x\,d a
- \int_{\rm bf} \rho_{\rm s}\,\bar g \cdot \delta x\,d a \nonumber\\
&+ \int_{\rm bf} (\delta U_{a_0} - T\,\delta S_{a_0} 
- \sum_i \mu_i\,\delta m_{i,a_0})\,d a_0 = 0, \label{bfvariational2}
\end{align}
for any variation such that the points of the body which belong to $\Gamma$ remain fixed. 

From this condition, we are now able to determine the set of the `local' (see below, after (\ref{surfaceinternalenergy1})) thermodynamic variables of state of the $\rm S_{bf}$ surface. The expression $\delta f_0 = \delta U_{a_0} - T\,\delta S_{a_0} - \sum_i \mu_i\,\delta m_{i,a_0}$ clearly depends on the variations of the thermodynamic variables of state of $\rm S_{bf}$, for a given point $x_0$ ($\in \rm S_{0,bf}$). What is shown in the preceding condition (\ref{bfvariational2}) is that $\int_{\rm bf} \delta f_0\,d a_0$, and then also $\delta f_0$ at a given point $x_0$, only depend on the vector field $\delta x$ defined on the $\rm S_{bf}$ surface, i.e., on the variation of only one geometrical variable: the field of the positions $x$ of the points of the surface. Since a thermodynamic variable of state is geometrically local, this geometrical variable must be, in fact, $(x(x_0),\nabla x(x_0),\nabla^2 x(x_0),...)$ at the point $x_0$ ($\nabla x$ is the covariant differential of $x$, on the surface $\rm S_{0,bf}$; $x_0 \in {\rm S_{0,bf}} \rightarrow x \in \rm S_{bf}$ is the restriction of $F_{0,\rm i}$ to $\rm S_{0,bf}$). Obviously, $x(x_0)$ cannot be a variable of state, because the thermodynamic state of the surface may remain unchanged under any translation, $x' = x + a$, whereas $(x(x_0),\nabla x(x_0),\nabla^2 x(x_0),...)$ is changed into $(x(x_0) + a,\nabla x(x_0),\nabla^2 x(x_0),...)$. The variable is then reduced to $(\nabla x(x_0),\nabla^2 x(x_0),...)$, i.e., at the first order, $\nabla x(x_0)$. Let $I_0$ be an isometry of ${\rm T}_x(\rm S_{bf})$ onto ${\rm T}_{x_0}(\rm S_{0,bf})$, such that $I_0 \cdot \nabla x(x_0)$ is a direct linear mapping on ${\rm T}_{x_0}(\rm S_{0,bf})$, and use the classical decomposition $I_0 \cdot \nabla x(x_0) = R_0(x_0) \cdot S(x_0)$, where $S(x_0)$ is a positive symmetric tensor and $R_0(x_0)$ a rotation in ${\rm T}_{x_0}(\rm S_{0,bf})$. With respect to another isometry $I_1 : {\rm T}_x({\rm S_{bf}}) \rightarrow  {\rm T}_{x_0}(\rm S_{0,bf})$, such that $I_1 \cdot \nabla x(x_0)$ is direct, the decomposition would be $I_1 \cdot \nabla x(x_0) = R_1(x_0) \cdot S(x_0)$, with $R_1(x_0) = R_{1,0} \cdot R_0(x_0)$ and $R_{1,0} = I_1 \cdot I^{-1}_0$, which shows that the symmetric tensor $S(x_0)$ only depends on $\nabla x(x_0)$ (and not on $I_0$). We thus have the unique decomposition $\nabla x(x_0) = I(x_0) \cdot S(x_0)$, where $I(x_0):{\rm T}_{x_0}({\rm S_{0,bf}}) \rightarrow {\rm T}_x(\rm S_{bf})$ is an isometry. In addition, under a rotation (of the whole space), $x \rightarrow x'$, $\nabla x(x_0)$ is changed into $\nabla x'(x_0) = R \cdot \nabla x(x_0)$ ($R$ being the associated vectorial rotation), but the symmetric tensor $S(x_0)$ remains unchanged, since $\nabla x'(x_0) = I'(x_0) \cdot S(x_0)$, with $I'(x_0) = R \cdot I(x_0)$. Since the thermodynamic state of the surface may remain unchanged under such an arbitrary rotation, whereas $\nabla x(x_0)$ is changed into $R \cdot \nabla x(x_0)$, but $S(x_0)$ remains unchanged, we may conclude that the true geometrical variable of state is $S(x_0)$. Other variables equivalent to $S = S(x_0)$ are $S^2$ (because $S$ is symmetric and positive) or the (Lagrangian) surface strain $e_{\rm s} = \frac{1}{2}(S^2 - I)$ ($I$ = identity operator), all these tensors being associated to the deformation of the surface (change of the scalar product), e.g., for $e_{\rm s}$:
\begin{gather}
e_{\rm s}(d x_0,d y_0) = \frac{1}{2}(d x \cdot d y - d x_0 \cdot d y_0),
\end{gather}
for arbitrary vectors $d x_0,d y_0 \in {\rm T}_{x_0}(\rm S_{0,bf})$. In conclusion, we may take $e_{\rm s}$ (at $x_0$) as the geometrical variable of state and, as mentioned above (as a consequence of (\ref{bfvariational2})), $\delta f_0 = \delta U_{a_0} - T\,\delta S_{a_0} - \sum_i \mu_i\,\delta m_{i,a_0}$ is then proportional to its variation $\delta e_{\rm s}$:
\begin{gather}
\delta U_{a_0} - T\,\delta S_{a_0} 
- \sum_i \mu_i\,\delta m_{i,a_0} = \pi_{\rm s} \cdot \delta e_{\rm s}\,. \label{surfacestress}
\end{gather}
The `coefficient' $\pi_{\rm s}$ of $\delta e_{\rm s}$ will be called the (Lagrangian) surface stress tensor at equilibrium, and $\pi_{\rm s} \cdot \delta e_{\rm s}$ the work of deformation of the surface, expressed per unit area in the reference state. Note that (\ref{surfacestress}) actually defines the symmetric part of the tensor $\pi_{\rm s}$, because $\delta e_{\rm s}$ is symmetric. Here, we consider that the surface stress ($\pi_{\rm s}$, and its Eulerian form $\sigma_{\rm s}$, below) is symmetric, but this may not be always the case (e.g., in the presence of surface couples of forces, acting on the interface; as for the case of volume stress: see the comment after (\ref{bequilibrium2})). In an equivalent form,
\begin{gather}
\delta U_{a_0} = T\,\delta S_{a_0} + \pi_{\rm s} \cdot \delta e_{\rm s}
+ \sum_i \mu_i\,\delta m_{i,a_0}. \label{surfaceinternalenergy1}
\end{gather}
Finally, this expression gives us the set of the `local' thermodynamic variables of state of the $\rm S_{bf}$ surface: $S_{a_0}$, $e_{\rm s}$ and $m_{i,a_0}\;(i=1,...,n)$. `Local' here means: for the subset of the thermodynamic states which can be reached by reversible transformations, from a given thermodynamic state. In such a subset of thermodynamic states, the surface stress $\pi_{\rm s}$ is then the partial derivative of $U_{a_0}$ with respect to the surface strain $e_{\rm s}$:
\begin{gather}
\pi_{\rm s}^{\alpha \beta} = \frac{\partial U_{a_0}}{\partial e_{\rm s, \alpha \beta}}\quad \textrm {with variables}\;(S_{a_0},e_{\rm s},m_{i,a_0})\label{surfacestress2}
\end{gather}
(components $\alpha, \beta \in \{1, 2\}$, with arbitrary coordinates on the surface; the two variables $e_{\rm s,12}$ and $e_{\rm s,21}$---although equal---being formally distinguished in (\ref{surfacestress2})).

If the above subset of thermodynamic states of the surface is equal to the set of all thermodynamic states, then the above `local' variables are the true (`global') thermodynamic variables of state of the surface. This occurs if any thermodynamic state can be reached by a reversible transformation from any other thermodynamic state. We know that such a property is valid for some volume bodies, such as viscoelastic solids or viscous fluids, for which any transformation becomes reversible when achieved at vanishing speed. We may assume that the surface of such bodies has the same property, in which case the above variables are the true thermodynamic variables of state of the surface.

From the definition of $\gamma_0$ and (\ref{surfacestress}), we may also write
\begin{gather}
\delta \gamma_0 = - S_{a_0}\,\delta T  + \pi_{\rm s} \cdot \delta e_{\rm s}
- \sum_i m_{i,a_0}\,\delta\mu_i,  \label{surfacegrandpotential}
\end{gather}
which gives another equivalent set of `local' thermodynamic variables of state: $T$, $e_{\rm s}$ and $\mu_i\;(i=1,...,n)$. The surface stress here appears as the partial derivative of $\gamma_0$ with respect to the surface strain:
\begin{gather}
\pi_{\rm s}^{\alpha \beta} = \frac{\partial \gamma_0}{\partial e_{\rm s, \alpha \beta}}\quad \textrm {with variables}\;(T,e_{\rm s},\mu_i)\label{surfacestress3}
\end{gather}
(a Shuttleworth type equation \cite{Shuttleworth:1950}; its Eulerian version is expressed in (\ref{surfacegrandpotentialeuler}), below).

As above, for the usual volume stress and strain tensors, we have a similar Eulerian representation for the surface stress and strain tensors ($\sigma_{\rm s}$ and $\delta \varepsilon_{\rm s}$, respectively; $\delta \varepsilon_{\rm s}(d x,d y) = \frac{1}{2} \delta (d x \cdot d y)$, for $d x,d y \in {\rm T}_x(\rm S_{bf})$), and the work of deformation of the surface:
\begin{gather}
\sigma_{\rm s} \cdot \delta \varepsilon_{\rm s}\,d a = \pi_{\rm s} \cdot \delta e_{\rm s}\,d a_0.\label{workdeformationsurface}
\end{gather}

Equation (\ref{surfaceinternalenergy1}) may also be written, with $d a_0$ fixed:
\begin{align}
\delta d U &= T\,\delta d S + \pi_{\rm s} \cdot \delta e_{\rm s}\,d a_0
+ \sum_i \mu_i\,\delta d m_i \nonumber\\
&= T\,\delta d S + \sigma_{\rm s} \cdot \delta \varepsilon_{\rm s}\,d a
+ \sum_i \mu_i\,\delta d m_i \label{surfaceinternalenergy2}
\end{align}
($d U = U_{a_0}\,d a_0$, etc; compare with (\ref{ff'internalenergy}) for a fluid--fluid surface, where $\sigma_{\rm s} = \gamma\,I$; $I$ is the identity operator on ${\rm T}_x(\rm S_{bf})$) or
\begin{gather}
\delta U_a = T\,\delta S_a  + (\sigma_{\rm s} - \gamma\,I) \cdot \delta \varepsilon_{\rm s}
+ \sum_i \mu_i\,\delta m_{i,a}
\end{gather}
(since $d U = U_a\,d a$, etc and $\delta d a = {\rm tr}(\delta \varepsilon_{\rm s})\,d a$); and with $d a_0$ variable:
\begin{gather}
\delta d U = T\,\delta d S + \pi_{\rm s} \cdot \delta e_{\rm s}\,d a_0
+ \gamma_0\,\delta d a_0 + \sum_i \mu_i\,\delta d m_i, \label{surfaceinternalenergy3}
\end{gather}
in which the surface stress $\pi_{\rm s}$ is associated with the deformation of a given element $ d a_0$, and the surface grand potential $\gamma_0$ with the creation of a new element of surface $\delta d a_0$. The Eulerian form of (\ref{surfacegrandpotential}) is, with $d a_0$ fixed:
\begin{gather}
\delta \gamma = - S_a\,\delta T  + (\sigma_{\rm s} - \gamma\,I) \cdot \delta \varepsilon_{\rm s}
- \sum_i m_{i,a}\,\delta\mu_i,  \label{surfacegrandpotentialeuler}
\end{gather}
since $\gamma\,d a = \gamma_0\,d a_0$ and $\delta d a = {\rm tr}(\delta \varepsilon_{\rm s})\,d a$ (compare with (\ref{gammaGibbs}) for a fluid--fluid surface).

Note that, although the three variables $e_{\rm s, \alpha \beta}$ are independent, it may occur, in some particular cases, that fewer variables are necessary. For example, if we a priori know that the tensor $\sigma_{\rm s}$ is isotropic (as for the isotropic pressure in a fluid, or the surface tension in a fluid--fluid surface), i.e., $\sigma_{{\rm s},\beta}^\alpha = \sigma_{\rm s}\, \delta_\beta^\alpha$ (eigenvalue also denoted $\sigma_{\rm s}$), then the expression (\ref{workdeformationsurface}) is reduced to $\sigma_{\rm s}\, \delta \varepsilon_{{\rm s},\alpha}^\alpha\,d a = \sigma_{\rm s}\,\delta d a$. In this case, we only need one variable, namely $d a$.

Moreover, note that the expression (\ref{workdeformationsurface}) obtained for the work of deformation of the surface, differs from that proposed in \cite{Nozieres-Wolf:1988,Muller-Saul:2004}:
\begin{gather}
(\bar{\sigma_t} \cdot \delta \varepsilon_t + \sigma_n \cdot \bar{\delta \varepsilon_n})\,d a, \label{NWMS}
\end{gather}
where $\sigma$ and $\delta \varepsilon$ are the volume stress and strain tensors, the subscripts $t$ and $n$ respectively refer to the tangential and normal components, and the symbol $\,\bar{}\,$ indicates an excess quantity on the surface. Since $\delta \varepsilon_{\rm s} = \delta \varepsilon_t$, the comparison of the two expressions shows that the surface stress $\sigma_{\rm s}$ here defined differs from its usual definition as the excess of tangential stress $\bar{\sigma_t}$. There are, in fact, too many variables in (\ref{NWMS}): 3 variables $\delta \varepsilon_t$ + 3 variables $\bar{\delta \varepsilon_n}$. The above thermodynamic approach showed that there are only 3 geometrical (`local') variables of state for the surface, namely the 3 components of $\delta e_{\rm s}$ (or $\delta \varepsilon_{\rm s}$). Note also that the continuity of the normal stress $\sigma_n = \sigma \cdot n$, when crossing the interface, is assumed in (\ref{NWMS}): however, the next section will show that there is generally a discontinuity of the normal stress, due to the presence of surface stress and gravity.

\section{Equations at surfaces and triple lines}

\subsection{Surfaces}

\noindent With (\ref{surfacestress}) and (\ref{workdeformationsurface}), the mechanical equilibrium condition (\ref{bfvariational2}) for the $\rm bf$ surface takes the form:
\begin{gather}
- \int_{\rm bf} (\sigma \cdot n) \cdot \delta x\,d a 
- \int_{\rm bf} p\,n \cdot \delta x\,d a
- \int_{\rm bf} \rho_{\rm s}\,\bar g \cdot \delta x\,d a
+ \int_{\rm bf} \sigma_{\rm s} \cdot \delta \varepsilon_{\rm s}\,d a = 0, \label{bfvariational3}
\end{gather}
for any variation such that $\delta x = 0$ on $\Gamma$. By application of Green's formula to the last term, this leads to the mechanical equilibrium equation of the surface:
\begin{gather}
{\rm div}\,\sigma_{\rm s} + \rho_{\rm s}\,\bar g + \sigma \cdot n + p\,n = 0. \label{bfequilibrium}
\end{gather}
With Cartesian coordinates $(x^i)\,(i=1,2,3)$ for the whole space and arbitrary curvilinear coordinates $(x^\alpha)\,(\alpha=1,2)$ for the surface (for $x \in \rm S_{bf}$, $(x^\alpha)$ and $(x^i)$ must be clearly distinguished), it may be written
\begin{gather}
\partial_\beta (\sigma_{\rm s}^{\alpha\beta} \partial_\alpha x^i) +
\Gamma_{\beta\gamma}^\beta \sigma_{\rm s}^{\alpha\gamma} \partial_\alpha x^i +
\rho_{\rm s} \bar g^i + \sigma^{ij} n_j + p n^i = 0,
\end{gather}
where $i,j \in \{1,2,3\}$ refer to space coordinates, $\alpha,\beta,\gamma \in \{1,2\}$ to surface coordinates, summation is performed on repeated indices, $\partial_\alpha = \displaystyle \frac{\partial}{\partial x^\alpha}$ and $\Gamma^\alpha_{\beta\gamma}$ are the Christoffel's symbols. A similar equation was obtained in a purely mechanical approach \cite{Gurtin-Murdoch:1975}. Note the similarity of this equation (\ref{bfequilibrium}) with the Cauchy volume one (\ref{bequilibrium1}).

Note also that (\ref{bfequilibrium}) may easily be generalized at the surface between two deformable bodies ($\rm b$ and $\rm b'$):
\begin{gather}
{\rm div}\,\sigma_{\rm s} + \rho_{\rm s}\,\bar g + \sigma \cdot n - \sigma' \cdot n = 0, \label{bb'equilibrium}
\end{gather}
where $\sigma'$ is the stress tensor in $\rm b'$, and $n$ is oriented from $\rm b'$ to $\rm b$.

Equation (\ref{bfequilibrium}) may be separated into tangential and normal components. The tangential component is
\begin{gather}
{\rm div_s}\,\sigma_{\rm s} + \rho_{\rm s}\,\bar g_t + (\sigma \cdot n)_t = 0, \label{bfequilibriumtangent}
\end{gather}
where the subscript $t$ indicates the vector component tangent to the surface and $\rm div_s$ is the surface divergence:
\begin{gather}
({\rm div_s}\,\sigma_{\rm s})^i = 
(\partial_\beta \sigma_{\rm s}^{\alpha\beta} 
+ \Gamma_{\gamma\beta}^\alpha \sigma_{\rm s}^{\gamma\beta}
+ \Gamma_{\gamma\beta}^\beta \sigma_{\rm s}^{\alpha\gamma})\partial_\alpha x^i.
\end{gather}
The normal component of (\ref{bfequilibrium}) is
\begin{gather}
\sigma_{\rm s} \cdot l_n + \rho_{\rm s}\,\bar g_n + \sigma_{nn} + p = 0, \label{bfequilibriumnormal}
\end{gather}
where $\bar g_n = \bar g \cdot n$, $\sigma_{nn} = (\sigma \cdot n)\cdot n = \sigma^{ij} n_i n_j$ and $l_n$ is the `fundamental form' 
\begin{gather}
l_{n,\alpha\beta} = (\partial_{\alpha\beta} x^i -
\Gamma_{\alpha\beta}^\gamma \partial_\gamma x^i)n_i,
\end{gather}
the eigenvalues of which are the principal curvatures, $\displaystyle\frac{1}{R_1}$ and $\displaystyle\frac{1}{R_2}$, of the surface (a curvature being considered positive, when its center is on the side of $n$). This equation (\ref{bfequilibriumnormal}) generalizes the classical
Laplace equation (\ref{pp'equilibrium}) for a fluid--fluid surface. Indeed, $\sigma_{nn} + p$ represents the difference of the normal pressures and, for isotropic surface stresses, $\displaystyle \sigma_{\rm s} \cdot l_n = \sigma_{\rm s}\,l_{n,\alpha}^\alpha 
= \sigma_{\rm s} (\frac{1}{R_1} + \frac{1}{R_2})$ (eigenvalue also denoted $\sigma_{\rm s}$).

Equation (\ref{bfequilibrium}) also shows that there is a discontinuity of the normal stress at the interface, the jump of which, $\sigma \cdot n + p\,n$, is due both to surface stress (term ${\rm div}\,\sigma_{\rm s}$, with components ${\rm div_s}\,\sigma_{\rm s}$ and $\sigma_{\rm s} \cdot l_n$) and to gravity (term $\rho_{\rm s}\,\bar g$).

\subsection{Triple lines}

\noindent Let us now consider a bounding surface $\Sigma$ which encloses two fluids ($\rm f$ and $\rm f'$) and the body $\rm b$, and apply the general equilibrium condition (\ref{bodyvariational1}), using the above expressions (\ref{divergencesigma}), (\ref{surfacestress}) and (\ref{workdeformationsurface}). Note that Green's formula applied to the work of deformation of the $\rm bf$ surface leads to the new term $- \int_{\rm bff'} (\sigma_{\rm s} \cdot \nu)\cdot w\,d l$ at the triple line, where $w = \delta x$ and $\nu$ is the unit vector normal to the $\rm bff'$ line, tangent to the $\rm bf$ surface, and oriented from the line to the interior of $\rm bf$ (here, $\sigma_{\rm s}$, $\nu$ and $w$ refer to the $\rm bf$ side). With the help of the equilibrium equations (\ref{bequilibrium1})--(\ref{bequilibrium2}) for $\rm b$, and (\ref{bfequilibrium}) for $\rm bf$ and $\rm bf'$, the general equilibrium condition is then reduced to:
\begin{align}
&- \int_{\rm bff'} ((\sigma_{\rm s,bf} \cdot \nu_{\rm bf})\cdot w_{\rm bf}
+ (\sigma_{\rm s,bf'} \cdot \nu_{\rm bf'})\cdot w_{\rm bf'})\,d l\nonumber\\
&- \int_{\rm bff'} \gamma_{\rm ff'}\,\nu_{\rm ff'}\cdot \delta X\,d l 
+ \int_{\rm bff'} (\gamma_{0,{\rm bf}} - \gamma_{0,{\rm bf'}}) \,\delta X_0\,d l_0 = 0, \label{bff'equilibrium1}
\end{align}
for any variation such that the two points of the $\rm bff'$ line which belong to $\Sigma$ remain fixed (both in space, and as points of the body).

\begin{figure}[htbp]
\begin{center}
\includegraphics[width=10cm]{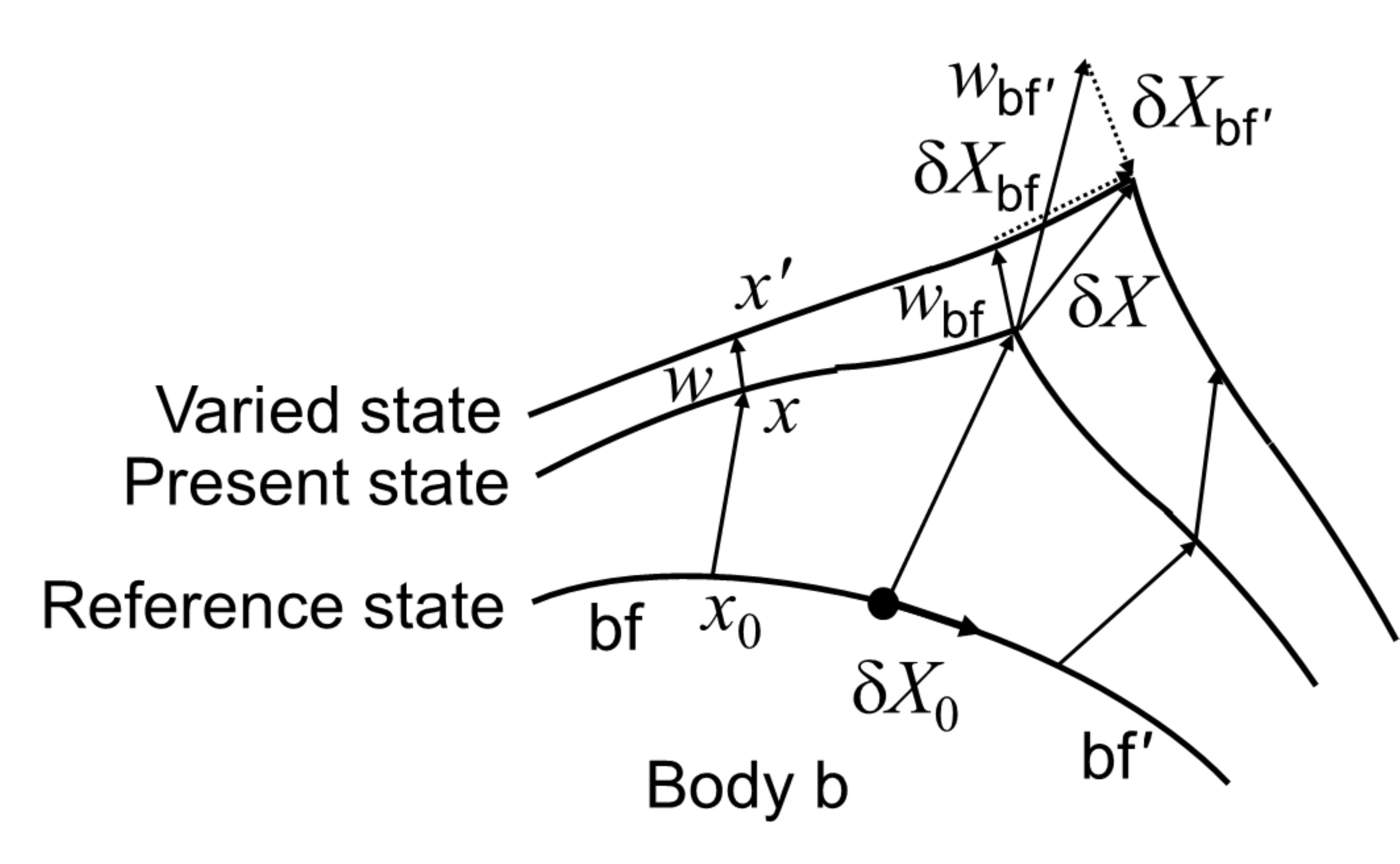}
\end{center}
\caption{Displacement $\delta X$ of the $\rm bff'$ triple line, expressed as $w_{\rm bf} + \delta X_{\rm bf}$ on the $\rm bf$ side, or $w_{\rm bf'} + \delta X_{\rm bf'}$ on the $\rm bf'$ side ($\delta X_{\rm bf}$ and $\delta X_{\rm bf'}$ are indicated by dotted arrows; see text).} \label{Displacements}
\end{figure}

The transformation $x_0 \rightarrow x$ restricted to the body--fluid surface, between the (ideal) reference and present states of the body, is continuous, but its differential $\nabla x$ is not continuous at the triple line, since the plane tangent to the $\rm bf$ surface and the plane tangent to the $\rm bf'$ surface (at the triple line) are different (figure~\ref{Displacements}). 
Its variation $w = \delta x$ will also be discontinuous at the triple line. From the $\rm bf$ side point of view, and if $\delta X_0 = 0$, the displacement of the $\rm bff'$ line would be equal to $w_{\rm bf}$. On the other hand, if $w = 0$ on $\rm bf$, and $\delta X_0 \neq 0$, this displacement would be $\delta X_{\rm bf} = \nabla x(x_0)_{\rm bf} \cdot \delta X_0$ ($x_0$ belongs to $\rm bff'$; $\delta X_0$ is here considered as a vector). In the general case, the displacement of the triple line will then be
\begin{align}
\delta X &= w_{\rm bf} + \delta X_{\rm bf} = w_{\rm bf} + \nabla x(x_0)_{\rm bf} \cdot \delta X_0\nonumber\\
&= w_{\rm bf'} + \delta X_{\rm bf'} = w_{\rm bf'} + \nabla x(x_0)_{\rm bf'} \cdot \delta X_0\label{displacementline}
\end{align}
(expression from the $\rm bf'$ side, in the second line; see figure~\ref{Displacements}; the component of this $\delta X$ normal to $\rm bff'$ is the $\delta X$ of (\ref{bff'equilibrium1}), but this has no effect on $\nu_{\rm ff'}\cdot \delta X$).

From (\ref{displacementline}) and the Eulerian form of the last term of (\ref{bff'equilibrium1}) ($\gamma\,d a = \gamma_0\,d a_0$; $\gamma$ is the excess of grand potential on the surface, per unit area in the present state), the equilibrium condition (\ref{bff'equilibrium1}) takes the form
\begin{align}
&- \int_{\rm bff'} (\sigma_{\rm s,bf} \cdot \nu_{\rm bf} 
+ \sigma_{\rm s,bf'} \cdot \nu_{\rm bf'} 
+ \gamma_{\rm ff'}\,\nu_{\rm ff'})\cdot \delta X\,d l\nonumber\\
&+ \int_{\rm bff'} ((\sigma_{\rm s,bf} \cdot \nu_{\rm bf})\cdot \delta X_{\rm bf} 
+ (\sigma_{\rm s,bf'} \cdot \nu_{\rm bf'})\cdot \delta X_{\rm bf'})\,d l\nonumber\\
&- \int_{\rm bff'} (\gamma_{\rm bf}\,\nu_{\rm bf} \cdot \delta X_{\rm bf}
+ \gamma_{\rm bf'}\,\nu_{\rm bf'} \cdot \delta X_{\rm bf'})\,d l = 0, \label{bff'equilibrium2}
\end{align}
which leads to two equilibrium equations at the triple line:
\begin{gather}
\sigma_{\rm s,bf} \cdot \nu_{\rm bf} 
+ \sigma_{\rm s,bf'} \cdot \nu_{\rm bf'} 
+ \gamma_{\rm ff'}\,\nu_{\rm ff'} = 0 \label{bff'equilibriumequ1}
\end{gather}
(by considering the triple line as fixed with respect to the body, i.e., $\delta X_0 = 0$, hence $\delta X_{\rm bf} = \delta X_{\rm bf'} = 0$) and
\begin{gather}
- \gamma_{\rm bf}\,\nu_{\rm bf} \cdot \delta X_{\rm bf}
- \gamma_{\rm bf'}\,\nu_{\rm bf'} \cdot \delta X_{\rm bf'}
+ (\sigma_{\rm s,bf} \cdot \nu_{\rm bf})\cdot \delta X_{\rm bf} 
+ (\sigma_{\rm s,bf'} \cdot \nu_{\rm bf'})\cdot \delta X_{\rm bf'} = 0. \label{bff'equilibriumequ2}
\end{gather}

Equation (\ref{bff'equilibriumequ1}) represents the equilibrium of the forces acting on the triple line, considered as fixed on the body ($\delta X_0 = 0$), whereas (\ref{bff'equilibriumequ2}) expresses the equilibrium relative to the motion of the triple line with respect to the body ($\delta X_0 \neq 0$). A similar situation was found in the case of the thin plate \cite{Olives:1993,Olives:1996}. Note that it was proposed \cite{Madasu-Cairncross:2004} that the equilibrium at the triple line involves not only the ($\rm bf$, $\rm bf'$ and $\rm ff'$) surface forces, but also a force which originates from the volume stresses $\sigma$ in the body and the singularity at the triple line. The obtained equation (\ref{bff'equilibriumequ1}) shows that there is no body volume contribution, and that the equilibrium only involves the forces exerted by the three surfaces ($\rm bf$, $\rm bf'$ and $\rm ff'$) on the triple line. This equation shows that surface stresses are forces acting on a line fixed on the body, and may easily be generalized to a line of contact between three deformable bodies ($\rm b$, $\rm b'$ and $\rm b''$):
\begin{gather}
\sigma_{\rm s,bb'} \cdot \nu_{\rm bb'} 
+ \sigma_{\rm s,b'b''} \cdot \nu_{\rm b'b''} 
+ \sigma_{\rm s,bb''} \cdot \nu_{\rm bb''} = 0 \label{bb'b"equilibriumequ1}
\end{gather}
(equilibrium of the forces acting on the triple line, considered as fixed with respect to the three bodies).

Let $\tau$ be a unit vector tangent to the triple line, $(\sigma_{{\rm bf},\nu\nu}$, $\sigma_{{\rm bf},\tau\nu})$ the components of $\sigma_{\rm s,bf} \cdot \nu_{\rm bf}$ in the basis $(\nu_{\rm bf},\tau)$, and use similar notations for $\sigma_{\rm s,bf'}$ with the basis $(\nu_{\rm bf'},\tau)$. Equation (\ref{bff'equilibriumequ1}) may then be separated into its normal and tangential components (with respect to the line):
\begin{gather}
\sigma_{{\rm bf},\nu\nu}\,\nu_{\rm bf} 
+ \sigma_{{\rm bf'},\nu\nu}\,\nu_{\rm bf'} 
+ \gamma_{\rm ff'}\,\nu_{\rm ff'} = 0 \label{bff'equilibriumequ1normal}\\
\sigma_{{\rm bf},\tau\nu} + \sigma_{{\rm bf'},\tau\nu} = 0. \label{bff'equilibriumequ1tangent}
\end{gather}

Equation (\ref{bff'equilibriumequ2}) may also be viewed as the expression of the equilibrium (\ref{bff'equilibrium2}), when $\delta X = 0$ and $\delta X_0 \neq 0$: this means that the triple line is fixed in space, but moves with respect to the body; in other words, the body moves with respect to the triple line, which remains fixed in space (the body then also moves with respect to its singularity, attached to the triple line). Since $\delta X = 0$ implies that $w_{\rm bf} = - \delta X_{\rm bf} \in {\rm T}_x(\rm S_{bf})$ and $w_{\rm bf'} = - \delta X_{\rm bf'} \in {\rm T}_x(\rm S_{bf'})$, (\ref{bff'equilibriumequ2}) may then be written as
\begin{gather}
- \gamma_{\rm bf}\,\nu_{\rm bf} \cdot \delta X_{\rm bf}
- \gamma_{\rm bf'}\,\nu_{\rm bf'} \cdot \delta X_{\rm bf'}
= (\sigma_{\rm s,bf} \cdot \nu_{\rm bf})\cdot w_{\rm bf} 
+ (\sigma_{\rm s,bf'} \cdot \nu_{\rm bf'})\cdot w_{\rm bf'}. \label{bff'equilibriumequ2bis}
\end{gather}
This equilibrium equation has a clear physical meaning: it states that the variation of surface energy (grand potential) due to the motion of the triple line with respect to the body (e.g., increase in bf area and decrease in bf' area) is equal to the work of the surface stresses acting on the triple line (although the triple line does not move in space, the points of the body situated at this line move by $w_{\rm bf}$ in the $\rm bf$ side, and by $w_{\rm bf'}$ in the $\rm bf'$ side).

From the expressions (\ref{displacementline}) of $\delta X_{\rm bf}$ and $\delta X_{\rm bf'}$, equation (\ref{bff'equilibriumequ2}):
\begin{gather}
((\sigma_{\rm s,bf} - \gamma_{\rm bf}\,I) \cdot \nu_{\rm bf}) \cdot \delta X_{\rm bf}
+ ((\sigma_{\rm s,bf'} - \gamma_{\rm bf'}\,I) \cdot \nu_{\rm bf'}) \cdot \delta X_{\rm bf'} = 0\label{bff'equilibriumequ3}
\end{gather}
may be written as
\begin{gather}
((\sigma_{\rm s,bf} - \gamma_{\rm bf}\,I) \cdot \nu_{\rm bf}) \cdot \nabla x(x_0)_{\rm bf}
+ ((\sigma_{\rm s,bf'} - \gamma_{\rm bf'}\,I) \cdot \nu_{\rm bf'}) \cdot \nabla x(x_0)_{\rm bf'} = 0 \label{bff'equilibriumequ4}
\end{gather}
(in this last equation, the covariant forms of the tensors $\sigma_{\rm s,bf}$ and $\sigma_{\rm s,bf'}$ are used, and $I$ denotes the covariant metric tensors on $\rm S_{bf}$ and $\rm S_{bf'}$, respectively).

In the reference state, let $\nu_0$ be the unit vector normal to the $\rm bff'$ line, tangent to the $\rm bf$ or $\rm bf'$ surface (in the reference state, there is no singularity at the triple line, and the planes tangent to $\rm S_{bf}$ and $\rm S_{bf'}$ are identical), and oriented from $\rm bf'$ to $\rm bf$, and $\tau_0$ the unit vector tangent to $\rm bff'$, with the same orientation as $\tau$. With the notations
\begin{align*}
\bar\nu_{\rm bf} &= \nabla x(x_0)_{\rm bf} \cdot \nu_0 
= a_{\nu\nu}\,\nu_{\rm bf} + a_{\tau\nu}\,\tau \\
\bar \nu_{\rm bf'} &= \nabla x(x_0)_{\rm bf'} \cdot \nu_0 
= -a'_{\nu\nu}\,\nu_{\rm bf'} + a'_{\tau\nu}\,\tau\\
\bar\tau &= \nabla x(x_0) \cdot \tau_0 = a_{\tau\tau}\,\tau
\end{align*}
($a_{\nu\nu}$, $a'_{\nu\nu}$, $a_{\tau\tau} > 0$), (\ref{bff'equilibriumequ3}) or (\ref{bff'equilibriumequ4}) is equivalent to the two equations
\begin{gather}
((\sigma_{\rm s,bf} - \gamma_{\rm bf}\,I) \cdot \nu_{\rm bf}) \cdot \bar\nu_{\rm bf} + ((\sigma_{\rm s,bf'} - \gamma_{\rm bf'}\,I) \cdot \nu_{\rm bf'}) \cdot \bar \nu_{\rm bf'} = 0 \label{bff'equilibriumequ5}\\
\sigma_{{\rm bf},\tau\nu} + \sigma_{{\rm bf'},\tau\nu} = 0.
\end{gather}
The second equation is the same as (\ref{bff'equilibriumequ1tangent}) (which shows that (\ref{bff'equilibriumequ3}) does not depend on the component of $\delta X_0$ along $\tau_0$), and the first one may be written as
\begin{gather}
(\sigma_{{\rm bf},\nu\nu} - \gamma_{\rm bf})\,a_{\nu\nu}
+ \sigma_{{\rm bf},\tau\nu}\,a_{\tau\nu}
- (\sigma_{{\rm bf'},\nu\nu} - \gamma_{\rm bf'})\,a'_{\nu\nu} 
+ \sigma_{{\rm bf'},\tau\nu}\,a'_{\tau\nu} = 0, \label{bff'equilibriumequ6}
\end{gather}
i.e.,
\begin{gather}
(\sigma_{{\rm bf},\nu\nu} - \gamma_{\rm bf})\,a_{\nu\nu}
- (\sigma_{{\rm bf'},\nu\nu} - \gamma_{\rm bf'})\,a'_{\nu\nu}
+ \sigma_{{\rm bf'},\tau\nu}(a'_{\tau\nu} - a_{\tau\nu}) = 0. \label{bff'equilibriumequ7}
\end{gather}
Another form of this equation is obtained after multiplication by $a_{\tau\tau}$:
\begin{gather}
(\sigma_{{\rm bf},\nu\nu} - \gamma_{\rm bf})(\frac{d a}{d a_0})_{\rm bf}
- (\sigma_{{\rm bf'},\nu\nu} - \gamma_{\rm bf'})(\frac{d a}{d a_0})_{\rm bf'} 
+ 2\sigma_{{\rm bf'},\tau\nu}(e'_{\tau\nu} - e_{\tau\nu}) = 0, \label{bff'equilibriumequ8}
\end{gather}
since
\begin{align*}
(\frac{d a}{d a_0})_{\rm bf} &= a_{\nu\nu}\,a_{\tau\tau}\\
(\frac{d a}{d a_0})_{\rm bf'} &= a'_{\nu\nu}\,a_{\tau\tau}\\
e_{\tau\nu} &= e_{\rm bf}(\tau_0,\nu_0) 
= \frac{1}{2}\bar\tau \cdot \bar\nu_{\rm bf} 
= \frac{1}{2} a_{\tau\nu}\,a_{\tau\tau}\\
e'_{\tau\nu} &= e_{\rm bf'}(\tau_0,\nu_0) 
= \frac{1}{2}\bar\tau \cdot \bar\nu_{\rm bf'} 
= \frac{1}{2} a'_{\tau\nu}\,a_{\tau\tau}.
\end{align*}

Let us show that the preceding equations do not depend on the reference state. Since
\begin{gather*}
\delta X_{\rm bf'} = \nabla x(x_0)_{\rm bf'} \cdot \delta X_0
= \nabla x(x_0)_{\rm bf'} \cdot (\nabla x(x_0)_{\rm bf})^{-1} \cdot \delta X_{\rm bf}
= \nabla x_{\rm r} \cdot \delta X_{\rm bf},
\end{gather*}
where
\begin{gather}
\nabla x_{\rm r} = \nabla x(x_0)_{\rm bf'} \cdot \nabla x(x_0)_{\rm bf}^{-1} \label{relativedifferentialtransformation}
\end{gather}
is the `relative' differential of the transformation, on the $\rm bf'$ side with respect to the $\rm bf$ side (this concept was defined in \cite{Olives-Bronner:1984}), the above equation (\ref{bff'equilibriumequ3}) takes the form
\begin{gather}
(\sigma_{\rm s,bf} - \gamma_{\rm bf}\,I) \cdot \nu_{\rm bf} + ((\sigma_{\rm s,bf'} - \gamma_{\rm bf'}\,I) \cdot \nu_{\rm bf'}) \cdot \nabla x_{\rm r} = 0 \label{bff'equilibriumequ9}
\end{gather}
(with the covariant forms of $\sigma_{\rm s,bf}$, $\sigma_{\rm s,bf'}$ and $I$, as above for (\ref{bff'equilibriumequ4})), which does not depend on the reference state. Indeed, if the previous reference state 0 (indicated by the subscript 0) is replaced by a new reference state 1 (subscript 1), such that $\nabla x_0(x_1)$ is continuous at the triple line, then
\begin{align*}
\nabla x(x_1)_{\rm bf} &= \nabla x(x_0)_{\rm bf} \cdot \nabla x_0(x_1)\\
\nabla x(x_1)_{\rm bf'} &= \nabla x(x_0)_{\rm bf'} \cdot \nabla x_0(x_1)
\end{align*}
(at the triple line), hence
\begin{align*}
\nabla x(x_1)_{\rm bf'} \cdot \nabla x(x_1)_{\rm bf}^{-1}
&= \nabla x(x_0)_{\rm bf'} \cdot \nabla x_0(x_1) 
\cdot \nabla x_0(x_1)^{-1} \cdot \nabla x(x_0)_{\rm bf}^{-1}\\
&= \nabla x(x_0)_{\rm bf'} \cdot \nabla x(x_0)_{\rm bf}^{-1},
\end{align*}
which shows that $\nabla x_{\rm r}$ does not depend on the reference state. Using the bases $(\nu_0,\tau_0)$, $(\nu_{\rm bf},\tau)$ and $(-\nu_{\rm bf'},\tau)$, (\ref{relativedifferentialtransformation}) may be written in the matrix form:
\begin{gather*}
\begin{pmatrix}
a_{{\rm r},\nu\nu} & 0 \\ a_{{\rm r},\tau\nu} & a_{{\rm r},\tau\tau}
\end{pmatrix}
= 
\begin{pmatrix}
a'_{\nu\nu} & 0 \\ a'_{\tau\nu} & a_{\tau\tau}
\end{pmatrix} 
\begin{pmatrix}
\frac{1}{a_{\nu\nu}} & 0 \\ -\frac{a_{\tau\nu}}{ a_{\nu\nu}\, a_{\tau\tau}} & \frac{1}{ a_{\tau\tau}}
\end{pmatrix}
= 
\begin{pmatrix}
\frac{a'_{\nu\nu}}{a_{\nu\nu}} & 0 \\ \frac{a'_{\tau\nu} - a_{\tau\nu}}{ a_{\nu\nu}} & 1 
\end{pmatrix},
\end{gather*}
which shows that
\begin{align*}
a_{{\rm r},\nu\nu} &= \frac{a'_{\nu\nu}}{a_{\nu\nu}}\\
a_{{\rm r},\tau\nu} &= \frac{a'_{\tau\nu} - a_{\tau\nu}}{ a_{\nu\nu}}\\
a_{{\rm r},\tau\tau} &= 1,
\end{align*}
and leads to another form of (\ref{bff'equilibriumequ9}) or (\ref{bff'equilibriumequ7}):
\begin{gather}
\sigma_{{\rm bf},\nu\nu} - \gamma_{\rm bf} 
- (\sigma_{{\rm bf'},\nu\nu} - \gamma_{\rm bf'})\,a_{{\rm r},\nu\nu} 
+ \sigma_{{\rm bf'},\tau\nu}\,a_{{\rm r},\tau\nu} = 0. \label{bff'equilibriumequ10}
\end{gather}

\subsection{Modified Young's equation}

\noindent If the body is a rigid solid, (\ref{bff'equilibriumequ1}) cannot be obtained from (\ref{bff'equilibrium2}), because it is not possible to move the triple line in the space ($\delta X \neq 0$) while this line remains fixed to the solid ($\delta X_0 = 0$). In fact, (\ref{bff'equilibrium2}) takes the form
\begin{gather}
- \int_{\rm bff'} (\gamma_{\rm ff'}\,\nu_{\rm ff'} + \gamma_{\rm bf}\,\nu_{\rm bf}
+ \gamma_{\rm bf'}\,\nu_{\rm bf'}) \cdot \delta X\,d l = 0
\end{gather}
($\delta X = \delta X_{\rm bf} = \delta X_{\rm bf'}$, since $w_{\rm bf} = w_{\rm bf'} = 0$), and leads to 
\begin{gather}
\gamma_{\rm bf} - \gamma_{\rm bf'} + \gamma_{\rm ff'}\,\cos \varphi_{\rm f} = 0 \label{Youngsequ}
\end{gather}
(since $\delta X \in {\rm T}_x(\rm S_{bf}) = {\rm T}_x(\rm S_{bf'})$ and $\nu_{\rm bf'} = -\nu_{\rm bf} $; $\varphi_{\rm f}$ is the angle of contact measured in the fluid $\rm f$), which is Young's classical equation. This shows that Young's classical equation only refers to the surface energies $\gamma$ (and not to the surface stresses $\sigma_{\rm s}$) and is related to the displacement of the triple line with respect to the solid.

Then, in the present case of a deformable body, (\ref{bff'equilibriumequ1}) cannot be compared with Young's classical equation (since it refers to the surface stresses and the triple line fixed on the body). The present generalization of Young's equation is in fact related to (\ref{bff'equilibriumequ2}) (or (\ref{bff'equilibriumequ3}), (\ref{bff'equilibriumequ4}), (\ref{bff'equilibriumequ7}), (\ref{bff'equilibriumequ9}), (\ref{bff'equilibriumequ10})), because this equation contains the surface energies $\gamma$ and is related to the displacement of the triple line with respect to the body. In order to eliminate the surface stresses $\sigma_{{\rm s},\nu\nu}$ in (\ref{bff'equilibriumequ10}), we use the projections of (\ref{bff'equilibriumequ1normal}) onto $\nu_{\rm bf}$ and $\nu_{\rm bf'}$:
\begin{gather*}
\sigma_{{\rm bf},\nu\nu} + \sigma_{{\rm bf'},\nu\nu}\,\cos\varphi_{\rm b}
+ \gamma_{\rm ff'}\,\cos\varphi_{\rm f} = 0\\
\sigma_{{\rm bf},\nu\nu}\,\cos\varphi_{\rm b} + \sigma_{{\rm bf'},\nu\nu}
+ \gamma_{\rm ff'}\,\cos\varphi_{\rm f'} = 0
\end{gather*}
($\varphi_{\rm f}$, $\varphi_{\rm f'}$ and $\varphi_{\rm b}$ are the three angles of contact, respectively measured in $\rm f$, $\rm f'$ and $\rm b$), which give
\begin{align}
\sigma_{{\rm bf},\nu\nu} &= \gamma_{\rm ff'}\,\frac{\sin\varphi_{\rm f'}}{\sin\varphi_{\rm b}}\\
\sigma_{{\rm bf'},\nu\nu} &= \gamma_{\rm ff'}\,\frac{\sin\varphi_{\rm f}}{\sin\varphi_{\rm b}}
\end{align}
(using $\varphi_{\rm b} + \varphi_{\rm f} + \varphi_{\rm f'} = 2\pi$). By introducing these expressions in (\ref{bff'equilibriumequ10}), we obtain the following modified form of Young's equation:
\begin{gather}
-\gamma_{\rm bf} + \gamma_{\rm bf'}\,a_{{\rm r},\nu\nu}
+ \gamma_{\rm ff'}\,\frac{\sin\varphi_{\rm f'} - a_{{\rm r},\nu\nu}\,\sin\varphi_{\rm f}}{\sin\varphi_{\rm b}}
+ \sigma_{{\rm bf'},\tau\nu}\,a_{{\rm r},\tau\nu} = 0,\label{modifYoungsequ1}
\end{gather}
or
\begin{gather}
-\gamma_{\rm bf} + \gamma_{\rm bf'}\,a_{{\rm r},\nu\nu}- \gamma_{\rm ff'}\,\cos\varphi_{\rm f} 
- \gamma_{\rm ff'}\,\sin\varphi_{\rm f}\,
\frac{\cos\varphi_{\rm b} + a_{{\rm r},\nu\nu}}{\sin\varphi_{\rm b}}
+ \sigma_{{\rm bf'},\tau\nu}\,a_{{\rm r},\tau\nu} = 0\label{modifYoungsequ2}
\end{gather}
(using $\sin\varphi_{\rm f'} = -\sin(\varphi_{\rm b} + \varphi_{\rm f})$). A modified form of Young's equation was previously presented in the case of the thin plate \cite{Olives:1993,Olives:1996}. Obviously, if $a_{{\rm r},\nu\nu} = 1$ and $a_{{\rm r},\tau\nu} = 0$ (or if $a_{{\rm r},\nu\nu} = 1$ and $\sigma_{{\rm bf'},\tau\nu} = 0$), this equation takes the form
\begin{gather*}
-\gamma_{\rm bf} + \gamma_{\rm bf'} - \gamma_{\rm ff'}\,\cos\varphi_{\rm f} 
- \gamma_{\rm ff'}\,\sin\varphi_{\rm f}\,
\frac{1 + \cos\varphi_{\rm b}}{\sin\varphi_{\rm b}} = 0,
\end{gather*}
which leads to the Young's classical equation (\ref{Youngsequ}) when $\varphi_{\rm b}$ tends to $\pi$.

\section{Conclusions}

\noindent This paper deals with the thermodynamic (and mechanical) equilibrium of a deformable body in contact with fluids, specially at the surfaces and the triple contact lines of the body. We have applied the `dividing surface' Gibbs approach \cite{Gibbs:1876-1878}, here refined with a new concept: the `ideal transformation' between two `ideal' states, which defines the `ideal' displacement of the material points, within the interface film, up to the dividing surface. The approach is based on a careful application of the general thermodynamic equilibrium criterion of Gibbs. The classical thermal and chemical equilibrium equations (\ref{Tequilibrium}), (\ref{muiequilibrium}), and the mechanical equations concerning the fluids (\ref{pequilibrium})--(\ref{gammanuequilibrium}) and the (volume part of the) body (\ref{bequilibrium1}), (\ref{bequilibrium2}) are first separated. We thus obtain a new equilibrium condition which only refers to the surfaces and the triple contact lines of the body. As a first consequence of this condition, it is shown that the `local' thermodynamic variables of state of the surface are only the temperature, the chemical potentials and the surface strain tensor (these are the true thermodynamic variables of state, if, e.g., the body is a viscoelastic solid or a viscous fluid). This leads to a new definition of the surface stress tensor (which differs from the usual definition as the excess of tangential stress) and to the corresponding surface thermodynamic
equations (\ref{surfaceinternalenergy1})--(\ref{surfacegrandpotentialeuler}), with the simple expression (\ref{workdeformationsurface}) for the work of deformation of the surface (which differs from the expression (\ref{NWMS}) of \cite{Nozieres-Wolf:1988,Muller-Saul:2004}). The mechanical equilibrium equation of the surface (\ref{bfequilibrium}) is then obtained. This equation, similar to the classical Cauchy one for the volume, may be separated into tangential (\ref{bfequilibriumtangent}) and normal (\ref{bfequilibriumnormal}) components. This normal component is in fact a generalization of the classical Laplace equation for a fluid--fluid surface. Finally, at the (body--fluid--fluid) triple contact lines, we show that there are two mechanical equilibrium equations. The first one (\ref{bff'equilibriumequ1}) (or (\ref{bff'equilibriumequ1normal}), (\ref{bff'equilibriumequ1tangent})) represents the equilibrium of the forces which act on the triple line, considered as a line fixed on the body (these forces are the surface stresses; there is no contribution from the volume stresses of the body, as proposed in \cite{Madasu-Cairncross:2004}). The very original second equation (\ref{bff'equilibriumequ3}) (or (\ref{bff'equilibriumequ4}), (\ref{bff'equilibriumequ7}), (\ref{bff'equilibriumequ9}), (\ref{bff'equilibriumequ10})) expresses the equilibrium relative to the motion of the triple line with respect to the body, while the line remains fixed in space: i.e., the body moves with respect to the triple line (and with respect to the singularity of the body, attached to this line), which remains fixed in space. This equation involves the surface energies, the surface stresses, and the `relative deformation' \cite{Olives-Bronner:1984} between the two sides of the body surface separated by the triple line. It leads to a strong modification of the Young's classical capillary equation, as shown in the equations (\ref{modifYoungsequ1}) or (\ref{modifYoungsequ2}). A similar situation was found in the case of the thin plate \cite{Olives:1993,Olives:1996}. An approximate solution of these equations, for an elastic solid, will be given in a future paper.

\section*{Acknowledgments}

\noindent We acknowledge financial support from CINaM-CNRS and ANR-08-NANO-036.

\end{document}